\documentclass[11pt]{article}

\newif\ifdraftfigures
\draftfiguresfalse
\newif\ifincludeappendixfigures
\includeappendixfigurestrue

\usepackage[final]{acl}

\usepackage{times}
\usepackage{latexsym}
\usepackage[T1]{fontenc}
\usepackage[utf8]{inputenc}
\usepackage{microtype}
\usepackage{inconsolata}

\ifdraftfigures
  \usepackage[draft]{graphicx}
\else
  \usepackage{graphicx}
\fi
\usepackage{booktabs}
\usepackage{tabularx}
\usepackage{multirow}
\usepackage{adjustbox}
\usepackage{amsmath}
\usepackage{amssymb}
\usepackage{array}
\usepackage{url}
\usepackage{xspace}
\usepackage{placeins}

\newcommand{\ToolRobustBench}{ToolRobustBench\xspace}
\newcolumntype{Y}{>{\raggedright\arraybackslash}X}

\title{\ToolRobustBench: Stage-Wise Perturbation Evaluation and Failure Diagnosis for Tool-Calling Agents}

\author{YiShan Zheng \and Yuan Wu \and Yi Chang}

\begin{document}
\maketitle

\begin{abstract}
Large language models (LLMs) rely on tool calling as a fundamental agent capability, enabling them to invoke external systems and complete tasks beyond text generation. However, clean end-to-end (E2E) success cannot identify where a tool-use failure originates or how it propagates through a call. We introduce \ToolRobustBench, a stage-wise diagnostic benchmark for tool-calling agents, where a tool-calling agent is an LLM system that selects a tool, supplies structured arguments, and interprets its returned feedback. \ToolRobustBench aligns four perturbation families with the tool-use pipeline: tool-interface, user-intent, tool-output/observation, and runtime-environment perturbations. It attributes failures to tool selection, schema grounding, argument binding, tool-output/runtime-feedback handling, and E2E task success. Experiments on 15,456 single-family instances across 7 models, 16 sampled local tools, 4 perturbation families, and 14 subtypes show high but non-uniform clean performance and substantial robustness degradation, with tool-output/observation perturbation the dominant bottleneck. Mixed-family experiments reveal non-additive failure patterns that are not explained by isolated single-family results. Thus, \ToolRobustBench provides a deterministic and cascade-aware benchmark for diagnosing robustness beyond clean tool-calling accuracy; the anonymized artifact is available at \url{https://anonymous.4open.science/r/toolrobustbench/}.
\end{abstract}

\section{Introduction}

Large language models (LLMs) increasingly rely on tool calling as a central mechanism for interacting with external systems. Unlike ordinary question answering, tool calling requires a sequence of decisions: inferring the user's intent, selecting an appropriate tool, binding the request to structured arguments, and interpreting returned results or execution feedback. Here, a \emph{tool-calling agent} is an LLM-based system performing these decisions, and the \emph{clean setting} is the unperturbed tool-calling condition. Each step can be disrupted in deployment. Tool names may be replaced with semi-opaque service identifiers, descriptions of neighboring tools may overlap, requests may be indirect or internally inconsistent, and returned feedback may contain schema drift, missing evidence, misleading candidates, or runtime faults. Evaluation based only on final accuracy in clean settings cannot show whether these disruptions are handled reliably or where a failed call first departs from the intended process.

Existing benchmarks cover large-scale tool use and stable API execution \citep{qin2024toolllm,guo2024stabletoolbench}, as well as interactive trajectories and noisy tool-use conditions \citep{lu2025toolsandbox,wang2026agentnoisebench}. However, they leave room for controlled evaluation that localizes the first failed stage of a noisy tool-calling process.

A failed tool call has both a symptom and a source. An incorrect-argument output, for example, may arise directly during argument binding or may follow from earlier tool-selection confusion induced by an ambiguous interface. Similarly, a runtime-failure symptom may result from corrupted arguments rather than from the execution environment. Recording only the final failure label conflates these cases: it can attribute an upstream error to the wrong capability and may incorrectly suggest that a perturbation crossed its intended boundary.

\ToolRobustBench{} addresses this gap through controlled, single-step evaluation with stage-aligned perturbations and cascade-aware attribution. It measures not only whether a model succeeds under noise, but where its behavior first fails and how the error propagates.

Our contributions are:

\begin{enumerate}
    \item \textbf{Stage-aligned perturbation taxonomy.} We define four perturbation families that enter at distinct points in the tool-calling pipeline: tool-interface, user-intent, tool-output/observation, and runtime-environment perturbations.
    \item \textbf{Cascade-aware error attribution.} We separate observed error labels from primary error labels, earliest failed stages, error source stages, and boundary-violation indicators, so that downstream symptoms are not treated as direct perturbation effects.
    \item \textbf{Deterministic and reproducible evaluation.} We evaluate sampled tools from a 40-tool local environment with stable gold arguments, expected results, and attribution signals, avoiding volatility from live APIs.
    \item \textbf{Evidence on isolated and combined robustness.} Across 15,456 single-family instances and representative mixed-family perturbations, we identify observation handling as the main bottleneck and show that mixed-family robustness is not captured by the weaker constituent family alone.
\end{enumerate}

\ToolRobustBench{} therefore complements existing tool-use benchmarks with reproducible, stage-localized explanations of robustness failures.

\begin{figure*}[t]
    \centering
    \includegraphics[width=0.95\linewidth]{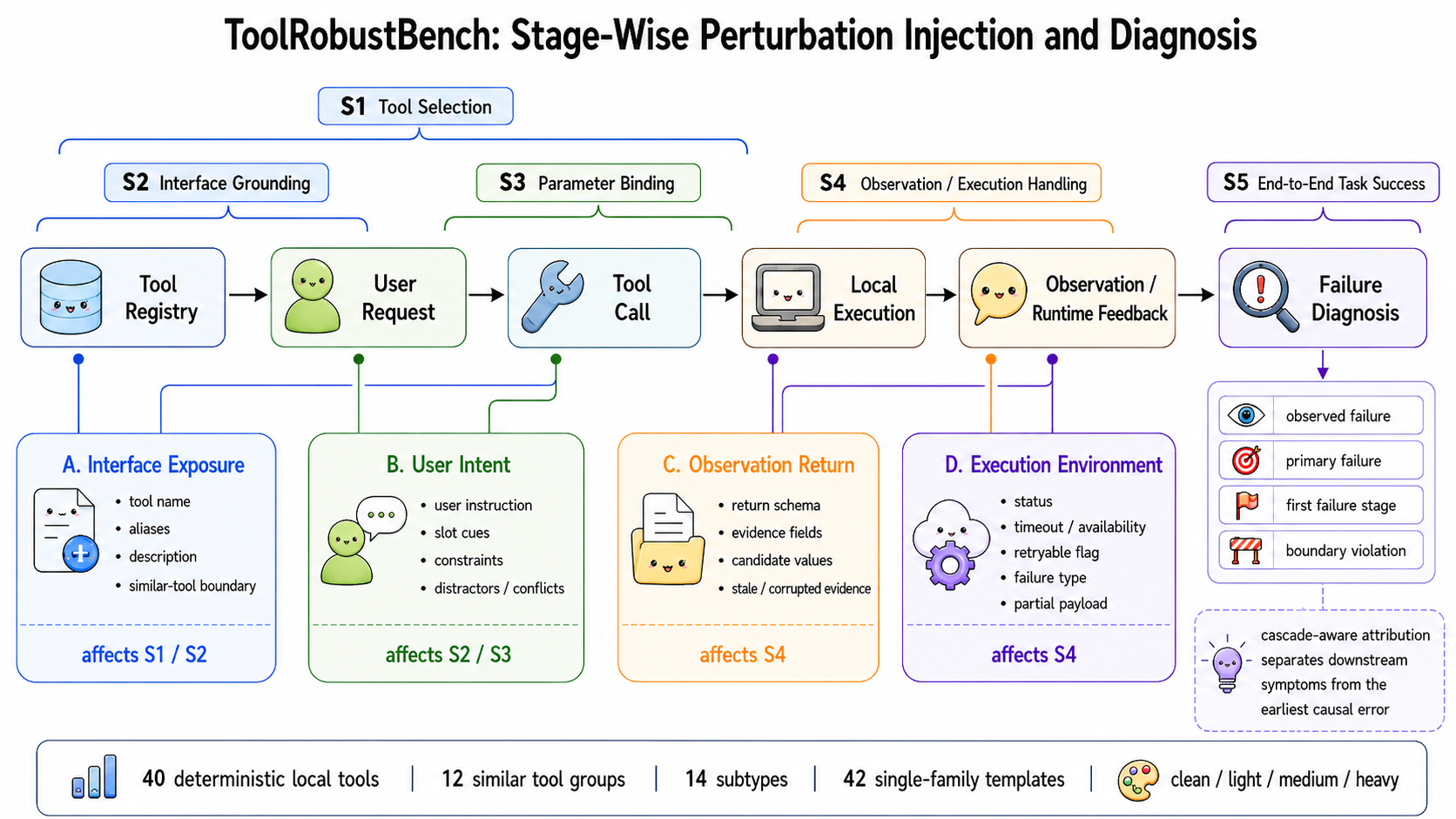}
    \caption{\ToolRobustBench{} diagnostic pipeline. The tool-interface perturbation (family A) alters the exposed registry before selection, the user-intent perturbation (family B) alters the request before grounding and binding, the tool-output/observation perturbation (family C) alters returned evidence after execution, and the runtime-environment perturbation (family D) alters execution state or feedback. S1--S5 locate the earliest failed stage.}
    \label{fig:pipeline}
\end{figure*}

\section{Related Work}

Evaluation has become a central problem for large language model research. Broad benchmarks such as MMLU \citep{hendrycks2021mmlu}, BIG-bench \citep{srivastava2023bigbench}, and HELM \citep{liang2023helm} establish capability and evaluation-protocol foundations, while survey work synthesizes this expanding landscape \citep{chang2024survey}. Tool-mediated action further makes process-level failure localization necessary.

\paragraph{Tool-calling evaluation.}
API-Bank \citep{li2023apibank} and ToolBench \citep{qin2024toolllm} extend evaluation from question answering to API retrieval, tool selection, and argument binding. Gorilla introduces APIBench for API selection and invocation under large and changing API collections \citep{patil2024gorilla}, while ToolAlpaca examines generalization to unseen tools using simulated tool-use cases \citep{tang2023toolalpaca}. StableToolBench \citep{guo2024stabletoolbench} addresses live-API instability with stable simulated APIs, and BFCL expands function-calling evaluation to serial, parallel, and agentic settings \citep{patil2025bfcl}. \ToolRobustBench{} adopts this reproducibility motivation while targeting controlled tool-interface, user-intent, tool-output/observation, and runtime-environment perturbations rather than unperturbed tool-calling performance alone.

\paragraph{Agent-process evaluation.}
AgentBench \citep{liu2024agentbench} and WebArena \citep{zhou2024webarena} evaluate LLM agents in interactive or web-based environments. ToolSandbox \citep{lu2025toolsandbox} and $\tau$-bench \citep{yao2025taubench} likewise focus on stateful tool use, for which evaluation depends on state updates, action sequences, and trajectory quality. \ToolRobustBench{} is complementary: its deterministic single-step tasks and post-execution decisions isolate observation recovery and runtime-state judgment without introducing long-horizon dependencies.

\paragraph{Noise-robustness evaluation.}
AgentNoiseBench \citep{wang2026agentnoisebench} categorizes realistic user-side and tool-side noise using solvability-preserving perturbations, while RoTBench \citep{ye2024rotbench} studies multi-level robustness for tool learning. \ToolRobustBench{} differs in providing stage-localized, cascade-aware diagnosis under controlled perturbations.

\paragraph{Failure attribution and agent safety.}
Who\&When \citep{zhang2025whowhen} is the closest prior work to our attribution motivation. It studies trajectory-level failure attribution in multi-agent systems and asks which agent and which step caused a task failure. \ToolRobustBench{} is complementary: it studies intra-call, pipeline-stage-level attribution inside a single tool-calling episode under controlled perturbations and deterministic scoring. AgentHarm \citep{andriushchenko2024agentharm} evaluates whether agents comply with harmful multi-step tasks; its safety focus differs from our robustness-diagnosis focus.

\begin{table*}[t]
\centering
\small
\begin{tabularx}{\linewidth}{l c c c c Y}
\toprule
Benchmark & Stage-aligned & Cascade-aware & Deterministic & Mixed-family & Main distinction \\
\midrule
\ToolRobustBench{} & Y & Y & Y & Y & Perturbation families are aligned to tool-call stages and failures are backtracked to primary sources. \\
RoTBench & Partial & N & N & N & Studies multi-level tool robustness but does not provide stage-level cascade attribution. \\
AgentNoiseBench & N & N & N & N & Covers realistic user/tool noise with solvability preservation, but not pipeline-stage attribution. \\
BFCL & N & N & N & N & Evaluates function-calling capability and formats without controlled perturbation diagnosis. \\
ToolSandbox / $\tau$-bench & N & N & N & N & Evaluate stateful or process-level tool use; do not isolate single-call perturbation entry points. \\
AgentHarm & N & N & N & N & Evaluates harmful agent behavior rather than robustness failure localization. \\
Who\&When & N & N & N & N & Attributes multi-agent trajectory failures, not intra-call tool-use perturbation failures. \\
\bottomrule
\end{tabularx}
\caption{Structured comparison with related benchmarks. Y/N/Partial indicate whether a benchmark directly supports the diagnostic property in its core protocol.}
\label{tab:related-comparison}
\end{table*}

\section{Design of \ToolRobustBench}

\subsection{Task Formulation}
\label{sec:task-formulation}

Each \ToolRobustBench{} instance is represented as
\begin{equation}
I=(x,\mathcal{R},t^*,a^*,o^*,y^*,p),
\end{equation}
where $x$ is the user request and $\mathcal{R}=\{t_1,\ldots,t_n\}$ is the candidate tool registry. Each tool $t\in\mathcal{R}$ exposes a name, a natural-language description, a JSON schema, a deterministic executor, and an output format. The gold target consists of the intended tool $t^*$, structured argument object $a^*$, deterministic tool output $o^*$, and expected final result $y^*$; $p$ records the perturbation family, subtype, and severity applied to the instance.

In pre-execution settings---clean, tool-interface, and user-intent---the model outputs a structured tool call $\hat{c}=(\hat{t},\hat{a})$. The scorer checks the selected tool, schema validity, bound arguments, and the result obtained by deterministic execution. In post-execution settings---tool-output/observation and runtime-environment---the model first completes the initial tool call and then, after receiving perturbed returned evidence or runtime feedback, outputs an observation-recovery or post-execution decision. These settings therefore evaluate not only selection and binding, but also interpretation of noisy evidence and execution-state feedback.

This is a full-process recovery setting within a single tool-calling episode, not a multi-turn recovery benchmark. For C/D families, the model can use the original request, the selected tool, the available clean context, and the perturbed returned evidence or runtime feedback to make its final decision. The benchmark therefore tests whether a model can recover from noisy observations or execution feedback inside one call; it does not claim to model persistent state repair across multiple user turns or chains of dependent tool calls.

At the instance level, success is binary: $s_i=1$ when the model selects $t^*$, produces a schema-valid argument object matching $a^*$, obtains the expected executed output where execution is applicable, and returns the expected final result $y^*$; otherwise $s_i=0$. For runtime-environment instances, success additionally requires the predicted runtime fields defined in Section~\ref{sec:taxonomy}---status, retryability, failure type, and final result---to agree with their gold targets whenever those judgments apply. Aggregate robustness metrics are defined separately below.

The diagnostic objective is not only to determine whether a tool call succeeds, but also to identify the earliest stage at which the model departs from the intended tool-use process. Accordingly, each evaluated record preserves the observed error label, primary error label, earliest failed stage, error source stage, cascade indicator, and boundary-violation indicator. To operationalize this diagnostic objective, we decompose each tool call into five capability axes, described next.

\subsection{Tool-Calling Capability Axes}

\ToolRobustBench{} uses five diagnostic capability axes, summarized in Table~\ref{tab:stages}: S1 tool selection, S2 schema grounding, S3 argument binding, S4 tool-output/runtime-feedback handling, and S5 end-to-end (E2E) task success. The axes roughly follow the temporal order of a tool call, but are used primarily for attribution: when a task fails, they identify the earliest source stage rather than every transition in a full execution trace.

\begin{table*}[t]
\centering
\small
\begin{tabularx}{\linewidth}{l l Y Y}
\toprule
Stage & Name & Target capability & Typical failures \\
\midrule
S1 & tool selection & Select the correct tool from candidate tools & no-call, tool misselection \\
S2 & schema grounding & Understand tool names, descriptions, and schemas & tool misselection, schema-format violation \\
S3 & argument binding & Bind user intent to arguments & missing argument, incorrect argument \\
S4 & tool-output/runtime-feedback handling & Interpret tool returns and runtime feedback & incorrect argument, schema-format violation, runtime failure \\
S5 & end-to-end task success & Whether the end-to-end task succeeds & all failure tags \\
\bottomrule
\end{tabularx}
\caption{The five diagnostic capability axes (S1--S5) in \ToolRobustBench. The axes locate the earliest source of a failed tool call; S5 reports final E2E success.}
\label{tab:stages}
\end{table*}

\subsection{Stage-Aligned Perturbation Taxonomy}
\label{sec:taxonomy}

\ToolRobustBench{} defines four single-family perturbation families aligned with these axes.

\textbf{Tool-interface perturbation} applies interface-level shifts, including surface changes to tool names, abstracted descriptions, and blurred boundaries among similar tools. It tests whether models rely excessively on lexical or descriptive cues.

\textbf{User-intent perturbation} varies the expression of the request through indirectness, underspecification, distracting context, and internal conflict. It tests whether models can recover the task goal and required arguments from natural-language input.

\textbf{Tool-output/observation perturbation} modifies returned content after the initial call. The model must recover the clean target from noisy feedback, distinguish the primary result from supporting evidence, and manage missing or conflicting evidence.

\textbf{Runtime-environment perturbation} enters during execution or the resulting runtime feedback. \emph{Runtime feedback} is the returned execution-state evidence made visible to the model; \emph{status} is its success, partial-success, or error state; \emph{failure type} categorizes an unsuccessful state; and \emph{final result} is the output that should be retained after interpreting all feedback. \emph{Retryability} denotes whether a failed or partial execution should be retried under the observed runtime state. For example, a timeout or temporary service unavailability may be retryable, whereas a permission denial or invalid argument error is usually non-retryable.

At the template level, a \emph{template} is a deterministic rule that derives a perturbed instance from a clean seed task. The single-family setting includes 14 subtypes with clean, light, medium, and heavy conditions. The main-text subtype definitions are summarized in Table~\ref{tab:subtypes}, with deterministic construction rules and intended diagnostic targets expanded in Appendix Table~\ref{tab:subtype-details}.

\subsection{Cascade-Aware Error Attribution}

Tool-calling errors often cascade across stages. A final incorrect-argument symptom may originate in S1/S2 tool misselection, and a runtime-failure symptom may be induced by upstream argument corruption rather than by the execution environment. To distinguish downstream symptoms from their earliest causes, \ToolRobustBench{} records six diagnostic fields during scoring.

\begin{table*}[t]
\centering
\small
\begin{tabularx}{\linewidth}{l Y Y}
\toprule
Field & Meaning & Purpose \\
\midrule
Observed error label & Final observed failure type & Characterizes end-to-end output behavior \\
Primary error label & Principal failure after cascade backtracking & Identifies the earliest substantive source \\
Earliest failed stage & First stage at which failure occurs & Localizes error onset to S1--S4 \\
Error source stage & Origin stage for a cascaded failure & Separates upstream causes from downstream symptoms \\
Cascade indicator & Whether an observed symptom is attributed to an earlier source & Quantifies failure propagation across stages \\
Boundary-violation indicator & Whether failure falls outside the expected family boundary & Audits perturbation construction and attribution \\
\bottomrule
\end{tabularx}
\caption{Cascade-aware diagnostic fields.}
\label{tab:diagnostic-fields}
\end{table*}

A \emph{cascade} occurs when the observed downstream symptom differs from the primary source recovered by backtracking; its \emph{cascade source} is the recorded error source stage. A \emph{boundary violation} occurs when the attributed source lies outside the stage boundary intended for that perturbation family. The main analysis focuses on primary failures, cascade rates, and boundary-violation indicators, reporting both the final symptom and whether it is consistent with the intended perturbation boundary.

Operationally, attribution is a deterministic backtracking procedure. The scorer first derives the observed error label through a fixed priority order: call presence, structured-format validity, tool-name resolution, canonical-tool match, required-argument presence, argument-value match, and execution or post-execution success. It then compares this observed label with the expected primary failures and allowed downstream spillovers for the perturbation family. If the observed label is an allowed spillover, the algorithm walks backward to the earliest compatible upstream stage and records that stage as the primary source. For example, in a tool-interface instance, a final incorrect-argument symptom can be reattributed to an upstream tool-misselection source if the model selected the wrong tool under perturbed interface names and then bound arguments for that wrong tool. A boundary violation is flagged only when the attributed source falls outside both the expected primary set and the allowed spillover set. No LLM judge is used in this attribution procedure.

\subsection{Metrics}

Let $s_i\in\{0,1\}$ be the instance-level success for model $m$, as defined above, and define its success on a set $\mathcal{D}$ as
\begin{align}
\mathrm{Succ}_m(\mathcal{D}) &=
\frac{1}{|\mathcal{D}|}\sum_{i\in \mathcal{D}}s_i.
\end{align}
Let $\mathcal{D}_{clean}$ be clean instances, $\mathcal{D}_{f,z}$ be instances in family $f\in\{A,B,C,D\}$ at severity $z$, and $Z=\{\text{light},\text{medium},\text{heavy}\}$. We report
\begin{align}
\mathrm{Clean}(m) &= \mathrm{Succ}_m(\mathcal{D}_{clean}), \\
\mathrm{Robust}_{f}(m) &= \frac{1}{3}\sum_{z\in Z}
\mathrm{Succ}_m(\mathcal{D}_{f,z}), \\
\mathrm{Overall}(m) &=
\frac{1}{4}\sum_{f\in\{A,B,C,D\}}\mathrm{Robust}_{f}(m), \\
\mathrm{Drop}_{f}(m) &=
\mathrm{Clean}(m)-\mathrm{Robust}_{f}(m).
\end{align}

For failed instances $\mathcal{F}_{f}$, let $e_i^{o}$ and $e_i^{p}$ be the observed and primary error labels and let $b_i$ indicate a boundary violation. Cascade and boundary-violation rates are
\begin{align}
\mathrm{CR}_{f} &=
\frac{1}{|\mathcal{F}_{f}|}\sum_{i\in\mathcal{F}_{f}}
\mathbf{1}[e_i^{o}\neq e_i^{p}], \\
\mathrm{BVR}_{f} &=
\frac{1}{|\mathcal{F}_{f}|}\sum_{i\in\mathcal{F}_{f}}
\mathbf{1}[b_i=1].
\end{align}
The primary-error distribution counts attributed source failures rather than downstream symptoms.

For a mixed-family pair $p=(f_1,f_2)$ at severity $z$, let $M(m,p,z)$ denote mixed-instance success, $W(m,p,z)$ the weaker constituent success, and $\bar{S}(m,p,z)$ their mean. We report
\begin{align}
\Delta(m,p,z) &= M(m,p,z)-W(m,p,z).
\end{align}
Figure~\ref{fig:cross-vs-single} reports $M$ and $W$; $\bar{S}$ is reported in the accompanying analysis.
A negative $\Delta$ indicates that the mixed-family perturbation is harder than its weaker isolated constituent, while a positive $\Delta$ does not imply improved robustness: it can arise from floor effects, more salient failure cues, explicit runtime feedback, or template-level variation between selected pairs and full family averages.

\section{Experimental Setup}

\subsection{Data Construction}
\label{sec:data-construction}

\paragraph{Construction overview.}
\ToolRobustBench{} is built by a custom construction pipeline rather than imported from an external benchmark or live API corpus. We first define a deterministic local-tool environment, create executable clean anchors, and then derive stage-aligned perturbed variants with attached gold and diagnostic targets. GPT is used only as an authoring aid for diversifying natural-language prompt templates before inclusion; the final task records, tool schemas, executor outputs, perturbations, and labels are produced by deterministic code. Appendix~\ref{app:data-construction} supplies a clean-seed example and template-level construction details.

\paragraph{Deterministic local-tool environment.}
\ToolRobustBench{} uses deterministic local tools whose outputs are fixed for valid argument objects. The environment contains 40 tools in 12 functional groups: weather/environment, conversion, retrieval, route/time, list operations, calendar, email, file, commerce, CRM/ticketing, database, and security. Each tool is an original local implementation with a hand-written JSON schema, parameter constraints, deterministic executor, and expected output format. The main experiments sample 16 tools from this environment using fixed seed 20260511. This design avoids live-API volatility and makes gold execution and scoring reproducible. Appendix Table~\ref{tab:tool-env} enumerates the complete environment.

\paragraph{Clean seed task construction.}
Clean seed tasks are produced by filling natural-language request templates with concrete argument values and gold tool calls. Specialized tools in the first functional groups use hand-written realistic scenarios with fixed argument sets, while generic business, file, database, and security tools use structured entity/scope/attribute slots to instantiate multiple anchors. Every candidate clean task is executed through \texttt{ToolExecutor}; only tasks whose gold tool and arguments resolve to the deterministic expected result are retained as perturbation anchors. A/B variants inherit the clean tool, arguments, and final result; C variants retain the clean target while altering returned evidence; and D variants may alter the expected status, retryability, failure type, or final result under the intended runtime state. Appendix Table~\ref{tab:clean-seed-example} gives one concrete seed record.

\paragraph{Perturbation family construction.}
Templates instantiate the four entry points defined in Section~\ref{sec:taxonomy}. Table~\ref{tab:subtypes} gives compact main-text definitions for all 14 subtypes. Detailed deterministic template operations and expected failure modes are provided in Appendix Table~\ref{tab:subtype-details}.

\begin{table}[t]
\centering
\scriptsize
\setlength{\tabcolsep}{3pt}
\begin{tabularx}{\columnwidth}{@{}p{0.24\columnwidth}X@{}}
\toprule
Family & Main-text subtypes \\
\midrule
A: Interface & A1 name shift; A2 description abstraction; A3 sibling confusability. \\
B: Intent & B1 indirectness; B2 underspecification; B3 distraction; B4 conflict. \\
C: Observation & C1 structure noise; C2 evidence loss; C3 corruption; C4 evidence competition. \\
D: Runtime & D1 availability failure; D2 variability; D3 feedback quality. \\
\bottomrule
\end{tabularx}
\caption{Main-text subtype map. Family A enters at the tool registry/schema, B at the user request, C at returned evidence, and D at runtime feedback; Appendix Table~\ref{tab:subtype-details} gives deterministic construction rules and expected failure modes.}
\label{tab:subtypes}
\end{table}

\paragraph{Severity-controlled expansion.}
For every non-clean subtype, a clean anchor is crossed with deterministic variants defined by family, subtype, severity, and operator template. The sampling rate controls how many clean anchors are drawn for each family--subtype--severity bucket, not the probability that an already selected instance is perturbed; within a perturbed condition, the designated perturbation is applied to every instance in that condition. The resulting dataset is stratified and paired: records in conditions such as A1-light or B4-heavy share clean anchors and differ only in the controlled perturbation applied to the target pipeline stage.

Severity is assigned through the operator catalog and a preflight gate. Each perturbation operator has an \texttt{intrinsic\_strength\_score} reflecting the degree of information degradation it introduces. For example, an A1 transparent alias such as replacing \texttt{get\_weather} with a readable alias receives a low score, while a semi-opaque or internal-code name receives a higher score. For user-intent, observation, and runtime templates, the static severity score is the sum of constituent operator scores; for interface templates, this operator score is combined with structural measurements such as name-surface change, description-token retention, alias removal, and sibling-tool similarity. Before model evaluation, the pipeline computes a \texttt{static\_strength\_score} for every subtype--severity template and enforces \texttt{light < medium < heavy}, subtype-purity, and nonzero-light checks. A failed check raises a preflight error rather than producing evaluation records.

The severity templates follow an additive stacking pattern. Light applies a single weak primary operator while keeping the target highly recoverable; medium applies a stronger primary operator or the light pattern plus an amplifier; heavy applies the strongest controlled operation, often by stacking the medium operator with one or more amplifiers while retaining a defined diagnostic target. Thus severity is a reproducible input-side stress axis. It makes the constructed perturbation strength monotonic, but it is not a guarantee that every model's empirical accuracy decreases monotonically within every subtype. Appendix Table~\ref{tab:severity-criteria} records the general criteria used for this expansion.

\paragraph{Diagnostic label construction.}
Constructed instances store the gold and attribution targets defined in Section~\ref{sec:task-formulation} and Section~\ref{sec:taxonomy}, with runtime-decision targets where applicable. Appendix Table~\ref{tab:construction-labels} maps these concepts to serialized record fields.

\paragraph{Dataset scale.}
The complete construction space contains 40 deterministic tools in 12 groups; the main experiment uses 16 sampled tools, four perturbation families, 14 subtypes, and conditions \{clean, light, medium, heavy\}. The raw single-family evaluation files were checked directly: each of seven models contributes 48 seed tasks for each of four family-specific clean baselines and $48 \times 14 \times 3$ perturbed records. Thus,
\[
7 \times \left[(4 \times 48) + (14 \times 3 \times 48)\right] = 15{,}456
\]
single-family raw model-output records, comprising 1,344 stored clean-baseline records and 14,112 perturbed records. Mixed-family records are constructed separately for representative mixed-family interactions and are not included in this single-family count.

\subsection{Data Quality and Scorer Reliability}

We audit 140 stratified single-family records, sampling ten records from each A1--D3 subtype and hiding automatic scorer labels during annotation. Human judgments show substantial agreement with automatic scoring on task success, observed error, primary error, and runtime-decision fields; a blind LLM annotation provides a secondary consistency check. Appendix~\ref{app:validity} and Appendix Table~\ref{tab:integrated-validation} detail the protocol and validation results.

\subsection{Models and Evaluation Scope}

We evaluate 7 models on the shared task set and sampled registry. For each model, the tool-interface and runtime-environment families contain 480 records each, whereas the tool-output/observation and user-intent families contain 624 records each; conditions are clean, light, medium, and heavy.

\subsection{Scoring Workflow}

Scoring follows the pre- and post-execution protocol in Section~\ref{sec:task-formulation}. Provider-specific parameters are used only for protocol compatibility; they do not alter task content or scoring criteria. Appendix~\ref{app:validity} describes scorer normalization and compatibility checks.

\section{Single-Family Results}

\subsection{Overall Robustness and Family Ranking}

\noindent\textbf{Finding 1: Clean tool-calling accuracy is high but non-uniform, and stage-wise robustness remains far from solved.}
The single-family experiment tests whether clean tool-calling performance predicts behavior under controlled perturbations. Table~\ref{tab:leaderboard} indicates that it does not: average clean-setting success is 0.979, with three models reaching 1.000, yet overall robustness under perturbation ranges from 0.766 to 0.664. Clean performance also contains meaningful task-level variation. Across 1,344 clean records, \texttt{filter\_items} reaches only 0.881, \texttt{query\_table} reaches 0.952, and \texttt{create\_support\_ticket} reaches 0.964, while \texttt{check\_permission}, \texttt{convert\_unit}, \texttt{get\_air\_quality}, \texttt{search\_policy}, and \texttt{sort\_items} reach 1.000. Model-level clean success ranges from 0.927 for \texttt{deepseek-v4-flash} to 1.000 for \texttt{claude-sonnet-4-6}, \texttt{deepseek-v4-pro}, and \texttt{qwen-plus}. The central result is therefore not a claim of global saturation, but the separation between mostly high clean capability and much weaker robustness under controlled perturbations.

\noindent\textbf{Finding 2: Tool-output/observation perturbation is the dominant bottleneck across all evaluated models.}
Family-level results identify where this separation arises. Averaged across models, tool-interface perturbation is least damaging at 0.918, followed by user-intent perturbation at 0.773 and runtime-environment perturbation at 0.688. Tool-output/observation perturbation is the bottleneck at 0.455, and no model reaches 0.60 in this family. In this evaluation, models handle tool selection and grounding more reliably than recovery from corrupted or incomplete returned evidence.

Individual profiles show the same pattern: strong interface robustness does not imply reliable observation recovery.

\begin{table*}[t]
\centering
\small
\begin{adjustbox}{width=\linewidth}
\begin{tabular}{lrrrrrrrl}
\toprule
Model & Clean & Overall & A Interface & B Intent & C Output & D Runtime & Heavy Macro & Main failure \\
\midrule
\texttt{claude-sonnet-4-6} & 1.000 & 0.766 & 0.968 & 0.785 & 0.568 & 0.745 & 0.583 & incorrect argument \\
\texttt{deepseek-v4-pro} & 1.000 & 0.749 & 0.926 & 0.809 & 0.512 & 0.748 & 0.516 & incorrect argument \\
\texttt{gpt-5.4-mini} & 0.984 & 0.712 & 0.856 & 0.781 & 0.503 & 0.708 & 0.469 & incorrect argument \\
\texttt{qwen-plus} & 1.000 & 0.706 & 0.931 & 0.802 & 0.380 & 0.713 & 0.507 & incorrect argument \\
\texttt{gemini-2.5-pro} & 0.979 & 0.692 & 0.900 & 0.748 & 0.481 & 0.637 & 0.454 & incorrect argument \\
\texttt{deepseek-v4-flash} & 0.927 & 0.669 & 0.958 & 0.783 & 0.319 & 0.616 & 0.475 & incorrect argument \\
\texttt{gemini-2.5-flash} & 0.964 & 0.664 & 0.889 & 0.700 & 0.420 & 0.648 & 0.436 & incorrect argument \\
\bottomrule
\end{tabular}
\end{adjustbox}
\caption{Main leaderboard recomputed from raw records. Tool-output/observation perturbation produces the largest drop for every model.}
\label{tab:leaderboard}
\end{table*}

\begin{figure}[t]
    \centering
    \includegraphics[width=\linewidth]{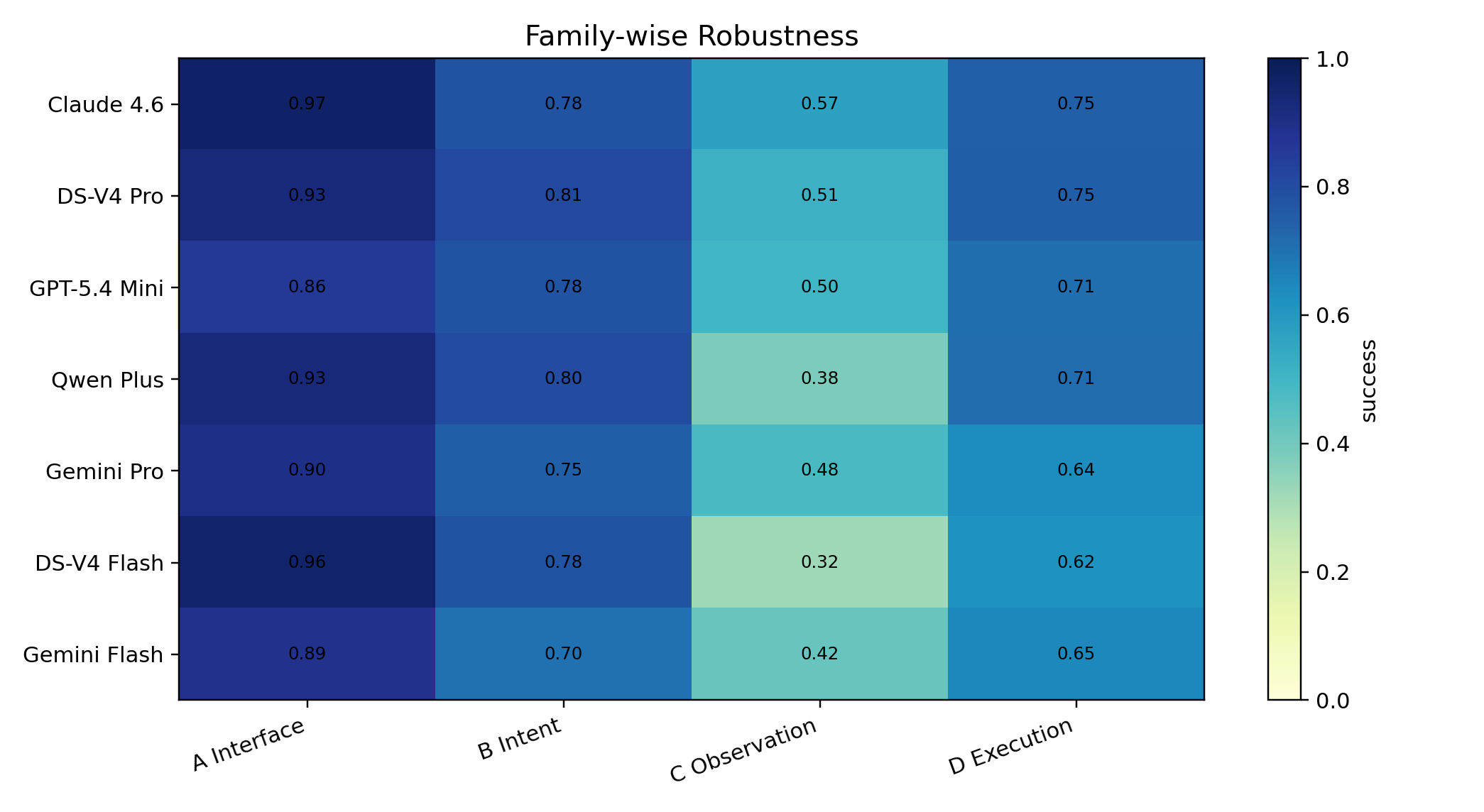}
    \caption{Family-level robustness heatmap. Tool-output/observation perturbation causes the largest drop for all models, and no model reaches 0.60 in family C.}
    \label{fig:family-heatmap}
\end{figure}

\subsection{Subtype and Severity Effects}

\noindent\textbf{Finding 3: Missing or competing returned evidence is substantially harder than surface-level interface change.}
Average success falls from 0.979 in clean conditions to 0.869, 0.724, and 0.491 under light, medium, and heavy perturbations, respectively, exposing failures largely hidden in clean evaluation.

Subtype results explain much of the family-level difference. C2 Return Evidence Loss is the hardest subtype (0.142), followed by C1 Return Structure Noise (0.460) and D2 Runtime Variability (0.469), whereas A2 Description Abstraction and A1 Name Surface Shift remain high at 0.927 and 0.930. Within families B and D, B3 Distraction and D2 Runtime Variability are the hardest subtypes, indicating sensitivity to competing context and unstable execution feedback.

Accuracy is not strictly monotonic with template severity for every subtype: explicit heavy runtime failures can be easier to classify than ambiguous stale-cache or partial-fragment combinations. Severity is therefore a controlled stress axis, not a guarantee of monotonic accuracy for each model and subtype.

\subsection{Recoverability of Evidence-Loss Perturbations}

C2 Return Evidence Loss is the hardest subtype, so we additionally validate whether its fixed gold targets remain recoverable from the perturbed information visible to the model. First, as an automated recovery proxy, we classify a C2 instance as \emph{provably solvable} if at least one evaluated model succeeds on it. Among 144 C2 instances, 49 are solved by at least one model. The solvable rates are 47.9\% for light, 52.1\% for medium, and 2.1\% for heavy C2, for an overall proxy-solvability rate of 34.0\%. Claude-sonnet-4-6 has the strongest C2 recovery behavior, with 27.8\% C2 success and 21 solo-rescue cases.

Second, we conduct a human recoverability audit on 60 C2 instances balanced across severities. The annotator judges whether the clean answer can be recovered from the perturbed tool return alone as YES, PARTIAL, or NO. The YES+PARTIAL rates are 100\% for light, 70\% for medium, and 35\% for heavy, with an overall rate of 68.3\%. These results show that C2 failures are not artifacts of systematically unrecoverable tasks. Rather, C2 exposes a gap between human evidence recovery and current model behavior, especially when returned fields or contextual evidence are missing.

\begin{table}[t]
\centering
\scriptsize
\begin{tabular}{lrr}
\toprule
Severity & Proxy solvable & Human recoverable \\
\midrule
Light & 23/48 (47.9\%) & 20/20 (100.0\%) \\
Medium & 25/48 (52.1\%) & 14/20 (70.0\%) \\
Heavy & 1/48 (2.1\%) & 7/20 (35.0\%) \\
\midrule
Total & 49/144 (34.0\%) & 41/60 (68.3\%) \\
\bottomrule
\end{tabular}
\caption{Recoverability validation for C2 evidence-loss perturbations. Proxy solvable means at least one evaluated model succeeds; human recoverable means YES or PARTIAL.}
\label{tab:c2-recoverability}
\end{table}

\subsection{Failure Taxonomy and Cascade Attribution}

\noindent\textbf{Finding 4: Cascade-aware attribution separates downstream symptoms from upstream failure sources.}
Across perturbed failures, \emph{incorrect argument} is the dominant primary error, followed by \emph{tool misselection} and \emph{runtime failure}; outright no-calls and schema-format violations are comparatively rare. Family composition is strongly concentrated: 89.1\% of tool-interface failures and 89.9\% of user-intent failures are tool misselections, 91.7\% of tool-output/observation failures are incorrect arguments, and 97.6\% of runtime-environment failures are runtime failures. An aggregate failure rate would collapse these distinct diagnostic modes.

\begin{figure}[t]
    \centering
    \includegraphics[width=\linewidth]{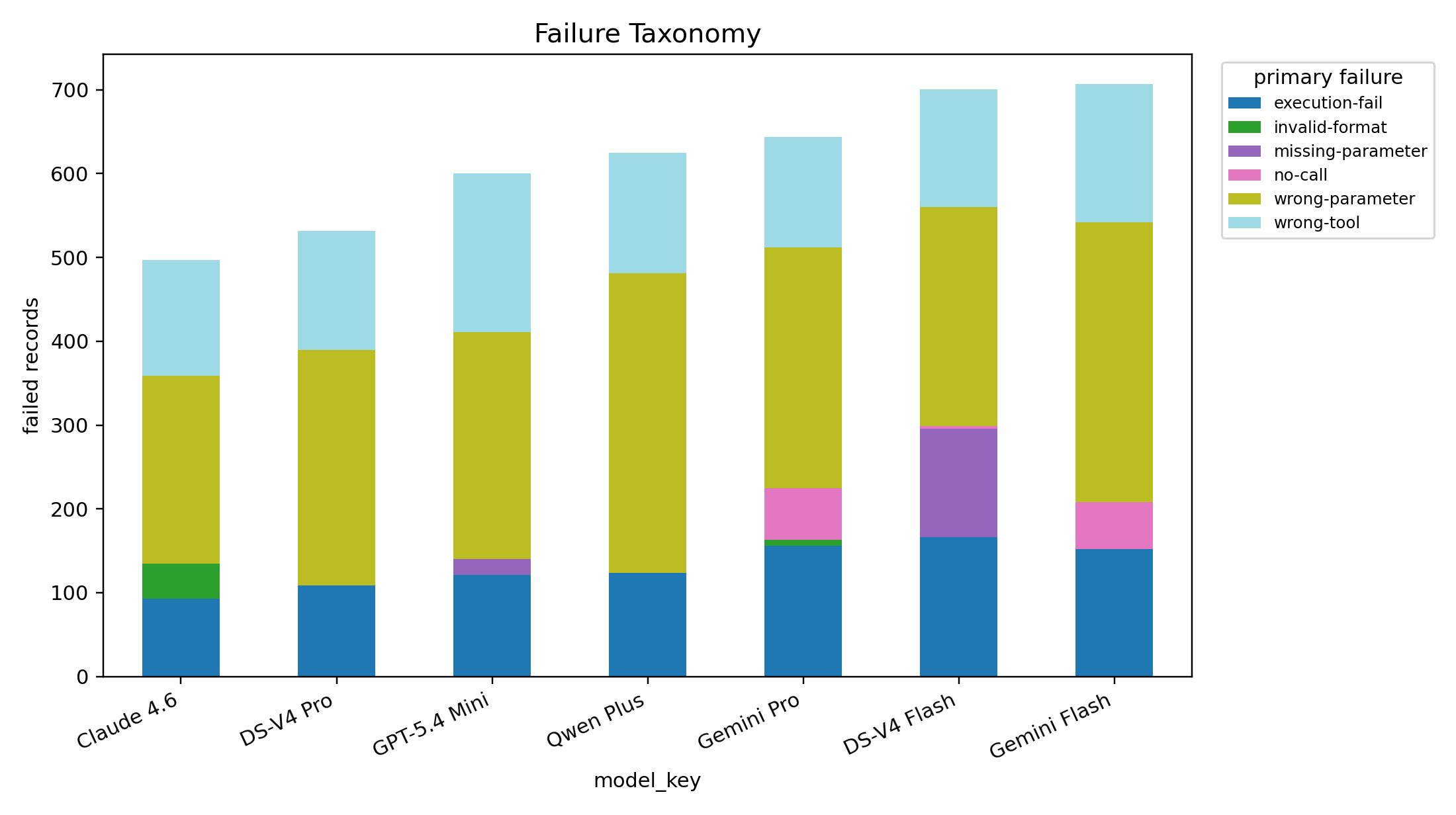}
    \caption{Primary failure taxonomy by perturbation family. Family-specific failure compositions show why aggregate success should be interpreted together with attribution.}
    \label{fig:failure-taxonomy}
\end{figure}

Cascade-aware attribution qualifies this interpretation. Runtime-environment perturbation has the highest cascade rate, 0.638, because runtime feedback occurs late in the tool chain and can expose upstream selection or argument errors. Tool-interface perturbation has a cascade rate of 0.352, often because boundary confusion produces downstream argument or execution symptoms. By contrast, tool-output/observation perturbation has a cascade rate of only 0.003, indicating that its failures usually arise directly during evidence selection or result recovery. Appendix Table~\ref{tab:boundary-stats} also distinguishes cascades from invalid construction: despite its high cascade rate, runtime-environment perturbation has a boundary-violation rate of only 0.006.

We further validate whether the attributed primary source is causally meaningful. In a counterfactual attribution study over 98 failed instances, we check whether correcting the output at the attributed primary error stage would recover the final task outcome. Recovery is 100\% for Tier-1 cases where the primary error is the only failing flag, 82.1\% for Tier-2 cascade cases where fixing the primary source removes downstream symptoms, and 0\% for Tier-3 multi-fault cases where several independent failures co-occur. Overall, 71/98 cases (72.4\%) recover under this counterfactual repair. By family, recovery is high for tool-interface (100.0\%), user-intent (81.0\%), and tool-output/observation (96.1\%), while runtime-environment recovery is 0.0\% because these instances require a second-stage multidimensional runtime judgment rather than a stage-1-only repair. A separate human audit on 30 deliberately ambiguous attribution cases obtains 74.2\% average agreement, with boundary-violation agreement at 93.3\%. These results support the use of cascade-aware labels as causal diagnostics rather than only descriptive tags.

\begin{table}[t]
\centering
\small
\begin{tabular}{lrr}
\toprule
Counterfactual tier & Count & Recovery \\
\midrule
Tier 1: single primary error & 48 & 48/48 (100.0\%) \\
Tier 2: cascade repair & 28 & 23/28 (82.1\%) \\
Tier 3: multi-fault & 22 & 0/22 (0.0\%) \\
\midrule
Total & 98 & 71/98 (72.4\%) \\
\bottomrule
\end{tabular}
\caption{Counterfactual validation of primary-error attribution. Recovery means that fixing the attributed primary failure would resolve the final task failure.}
\label{tab:counterfactual-attribution}
\end{table}

\subsection{Post-Execution Diagnostics}

Runtime-environment instances require a post-execution decision about status, retryability, failure type, and final result, as defined in Section~\ref{sec:taxonomy}. Averaged across models, matching is highest for retryability (0.949), followed by failure type (0.917), status (0.783), and result (0.738). Models therefore identify many explicit runtime labels but are less reliable at recovering the final result from mixed feedback. Appendix Figures~\ref{fig:app-stage2} and \ref{fig:app-gemini-profile} provide detailed profiles.

\subsection{Diagnostic Labels Guide Targeted Interventions}

The diagnostic labels are also actionable: they indicate which intervention should or should not help. For B-family user-intent perturbations, the diagnostic profile shows that many apparent S3 argument-binding errors are downstream symptoms of S1 tool-selection failures. Consistent with that diagnosis, adding chain-of-thought instructions for parameter binding does not improve \texttt{gpt-5.4-mini} on 30 B-family tasks: E2E success remains 21/30 (70.0\%) before and after the S3-targeted intervention. This negative result is useful because it shows that an S3 fix cannot repair S1-rooted failures.

Conversely, for A-family tool-interface perturbations, the diagnostic profile points to S1 tool-selection errors induced by perturbed tool names and descriptions. Providing explicit aliases next to the perturbed tool names raises \texttt{gpt-5.4-mini} success from 21/30 (70.0\%) to 30/30 (100.0\%). Together, these two intervention studies show that the labels can both reject a mismatched intervention and prescribe an effective one.

\begin{table}[t]
\centering
\scriptsize
\begin{tabularx}{\columnwidth}{p{0.28\columnwidth} c c X}
\toprule
Intervention & Baseline & Result & Diagnostic implication \\
\midrule
B-family S3 CoT & 21/30 & 21/30 & S3 prompting does not fix S1-rooted tool-selection failure. \\
A-family S1 aliases & 21/30 & 30/30 & Interface clarification fixes S1-driven misselection. \\
\bottomrule
\end{tabularx}
\caption{Targeted intervention studies guided by diagnostic labels on \texttt{gpt-5.4-mini}.}
\label{tab:diagnostic-interventions}
\end{table}

\section{Mixed-Family Results}
\label{sec:mixed}

We evaluate five interpretable subtype pairs at medium and heavy severity: A2+B3, A3+C4, A1+D1, B4+C4, and B2+D2. They cover distinct stage interactions, include observation-related combinations because family C is the dominant bottleneck, and retain non-degenerate accuracy ranges for model comparison.

We exclude degenerate mixtures that approach all-zero performance and combinations without an interpretable additional mechanism. In particular, many C--D mixtures are incompatible: C assumes an existing observation to corrupt, whereas D may replace that return with error or partial-execution feedback, preventing clean attribution. The broader \texttt{gpt-5.4-mini} composition matrix is reported in Appendix Figure~\ref{fig:app-gpt-cross-matrix}.

\begin{figure}[t]
    \centering
    \includegraphics[width=\linewidth]{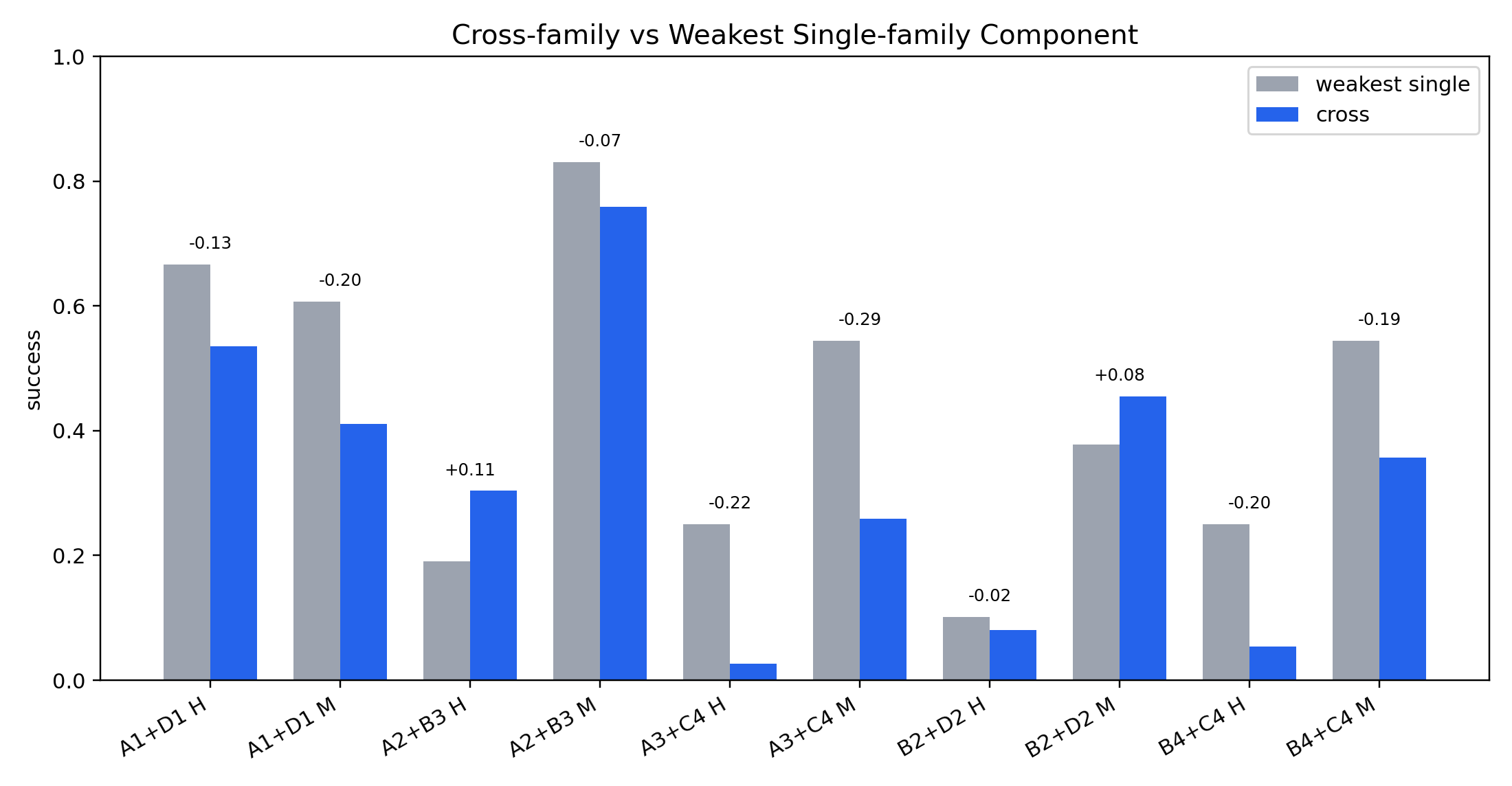}
    \caption{Selected mixed-family success compared with the weaker constituent baseline. Negative $\Delta$ indicates additional measured degradation under composition.}
    \label{fig:cross-vs-single}
\end{figure}

Each model--combination--severity cell contains 16 instances, yielding 70 comparable cells. Figure~\ref{fig:cross-vs-single} compares mixed success with its weaker isolated constituent baseline.

\noindent\textbf{Finding 5: Mixed-family perturbations often introduce negative interactions not predictable from isolated single-family robustness.}
Mixed-family success is generally lower than its single-family comparators. At medium severity, mean success across the five pairs and seven models is 0.448, compared with 0.621 for the weaker constituent and 0.754 for the mean of both constituents. At heavy severity, mixed-family success decreases to 0.200, compared with 0.292 and 0.507 for these two baselines. Thus, combined difficulty is not generally approximated by taking the minimum of two isolated scores. Relative to the mean of both constituents, the mixed-family gaps are -0.306 at medium severity and -0.307 at heavy severity.

The strongest negative interactions occur for A3+C4 and B4+C4. A3+C4 combines inter-tool confusability with competing returned evidence, while B4+C4 combines intent conflict with evidence competition. Both pairs push failures toward incorrect arguments, showing that observation noise can amplify earlier selection or intent uncertainty. Detailed pair-level failure compositions are reported in Appendix Figures~\ref{fig:app-cross-failure} and~\ref{fig:app-cross-model-sheet}.

Not every pair incurs additional measured loss. A2+B3 at heavy severity and B2+D2 at medium severity yield positive deltas ($+0.114$ and $+0.077$). These are apparent reversals rather than robustness improvements: floor effects or more salient feedback can make a composed condition easier to classify.

Taken together, these results show that mixed-family difficulty cannot be reduced to the weaker single-family component. Combinations involving tool-output/observation perturbations are particularly damaging because earlier interface or intent uncertainty is followed by unreliable evidence selection. Appendix Figures~\ref{fig:app-cross-success}, \ref{fig:app-cross-failure}, and \ref{fig:app-cross-model-sheet} report selected-pair success and failure diagnostics; Appendix Figure~\ref{fig:app-gpt-cross-matrix} reports the broader \texttt{gpt-5.4-mini} full mixed-family composition matrix.

\section{Discussion}

These results suggest that tool-use evaluation should combine clean success, stage-localized perturbations, and cascade-aware attribution. High clean accuracy hides failures in observation recovery and runtime-state judgment, while observed failure labels alone may point to the wrong corrective action when their sources lie upstream. Mixed-family results further show that deployed agents should be tested under targeted cross-stage interactions rather than only isolated perturbations. Interfaces should expose provenance, freshness, and machine-decidable runtime status to reduce observation and feedback ambiguity.

The validation studies sharpen this interpretation. C2 recoverability results show that evidence-loss failures are not merely artifacts of unrecoverable targets: humans recover at least partial answers in 68.3\% of sampled C2 cases, while the best model succeeds on only 27.8\%. Counterfactual attribution shows that primary-error labels recover 72.4\% of failed instances when repaired at the attributed stage, and targeted interventions show that the labels can distinguish useful S1 repairs from ineffective S3 prompting. Thus, the benchmark is not only a ranking instrument; it supplies a triage signal for improving tool descriptions, prompt design, observation schemas, and runtime feedback.

\section{Conclusion}

\ToolRobustBench{} is a stage-aligned, deterministic, and cascade-aware benchmark for tool-calling agents. Across 15,456 single-family instances and representative mixed-family perturbations, robustness drops sharply despite high clean success, especially under tool-output/observation perturbation. Recoverability, counterfactual attribution, targeted intervention, and human-audit studies support the validity and utility of the diagnostic labels. Mixed-family robustness cannot be inferred from the weaker constituent alone. The benchmark identifies where failures begin and which interface or runtime mechanisms merit improvement.

\section{Limitations}

The benchmark is intentionally controlled. Although the single-family results cover 14 subtypes, the main text reports five representative mixed-family combinations rather than an exhaustive ranking of pairwise templates: degenerate all-zero mixtures do not compare models, and some C--D mixtures lack a well-defined observation target once execution state is changed. Deterministic local tools improve reproducibility but cannot represent all behaviors of live API ecosystems. We expect the measurement principle that clean success does not predict perturbation robustness, the importance of observation/evidence recovery, and the value of stage-aligned attribution to transfer broadly to live APIs when wrappers expose comparable tool-call, argument, observation, and runtime-state fields. By contrast, absolute success rates, exact cascade rates, and D-family runtime conclusions are more specific to our controlled local setting; production APIs add authentication, rate limits, stateful side effects, vendor-specific error semantics, and long-horizon dependencies. The single-step tasks do not test long-horizon planning, persistent state, or dependencies across tools. Models are accessed through different providers or OpenAI-compatible backends, leaving protocol compatibility as a residual source of variation. Our human-validation sample is stratified for subtype coverage and deliberately includes many heavy and failed cases; consequently, it is appropriate for validating label and scorer behavior, but should not be interpreted as an unbiased estimate of the prevalence of ambiguous or difficult cases in the full benchmark. The annotation protocol also permits conservative \texttt{unclear} judgments when visible evidence does not uniquely determine a label, reducing the effective sample size for some agreement measures. Finally, our primary results are recomputed from raw records; Appendix~\ref{app:validity} documents human auditing, scorer normalization, and compatibility checks.

\bibliography{main}

@inproceedings{li2023apibank,
  title = {{API}-Bank: A Comprehensive Benchmark for Tool-Augmented {LLM}s},
  author = {Li, Minghao and Zhao, Yingxiu and Yu, Bowen and Song, Feifan and Li, Hangyu and Yu, Haiyang and Li, Zhoujun and Huang, Fei and Li, Yongbin},
  booktitle = {Proceedings of the 2023 Conference on Empirical Methods in Natural Language Processing},
  year = {2023},
  address = {Singapore},
  publisher = {Association for Computational Linguistics},
  pages = {3102--3116},
  doi = {10.18653/v1/2023.emnlp-main.187},
  url = {https://aclanthology.org/2023.emnlp-main.187/}
}

@article{chang2024survey,
  title = {A Survey on Evaluation of Large Language Models},
  author = {Chang, Yupeng and Wang, Xu and Wang, Jindong and Wu, Yuan and Yang, Linyi and Zhu, Kaijie and Chen, Hao and Yi, Xiaoyuan and Wang, Cunxiang and Wang, Yidong and Ye, Wei and Zhang, Yue and Chang, Yi and Yu, Philip S. and Yang, Qiang and Xie, Xing},
  journal = {ACM Transactions on Intelligent Systems and Technology},
  volume = {15},
  number = {3},
  pages = {1--45},
  year = {2024},
  publisher = {Association for Computing Machinery},
  doi = {10.1145/3641289},
  url = {https://doi.org/10.1145/3641289}
}

@inproceedings{qin2024toolllm,
  title = {{ToolLLM}: Facilitating Large Language Models to Master 16000+ Real-world {API}s},
  author = {Qin, Yujia and Liang, Shihao and Ye, Yining and Zhu, Kunlun and Yan, Lan and Lu, Yaxi and Lin, Yankai and Cong, Xin and Tang, Xiangru and Qian, Bill and Zhao, Sihan and Hong, Lauren and Tian, Runchu and Xie, Ruobing and Zhou, Jie and Gerstein, Mark and Li, Dahai and Liu, Zhiyuan and Sun, Maosong},
  booktitle = {The Twelfth International Conference on Learning Representations},
  year = {2024},
  url = {https://openreview.net/forum?id=dHng2O0Jjr}
}

@inproceedings{guo2024stabletoolbench,
  title = {{StableToolBench}: Towards Stable Large-Scale Benchmarking on Tool Learning of Large Language Models},
  author = {Guo, Zhicheng and Cheng, Sijie and Wang, Hao and Liang, Shihao and Qin, Yujia and Li, Peng and Liu, Zhiyuan and Sun, Maosong and Liu, Yang},
  booktitle = {Findings of the Association for Computational Linguistics: ACL 2024},
  year = {2024},
  address = {Bangkok, Thailand},
  publisher = {Association for Computational Linguistics},
  pages = {11143--11156},
  doi = {10.18653/v1/2024.findings-acl.664},
  url = {https://aclanthology.org/2024.findings-acl.664/}
}

@inproceedings{lu2025toolsandbox,
  title = {{ToolSandbox}: A Stateful, Conversational, Interactive Evaluation Benchmark for {LLM} Tool Use Capabilities},
  author = {Lu, Jiarui and Holleis, Thomas and Zhang, Yizhe and Aumayer, Bernhard and Nan, Feng and Bai, Haoping and Ma, Shuang and Ma, Shen and Li, Mengyu and Yin, Guoli and Wang, Zirui and Pang, Ruoming},
  booktitle = {Findings of the Association for Computational Linguistics: NAACL 2025},
  year = {2025},
  address = {Albuquerque, New Mexico},
  publisher = {Association for Computational Linguistics},
  pages = {1160--1183},
  doi = {10.18653/v1/2025.findings-naacl.65},
  url = {https://aclanthology.org/2025.findings-naacl.65/}
}

@inproceedings{yao2025taubench,
  title = {$\tau$-bench: A Benchmark for Tool-Agent-User Interaction in Real-World Domains},
  author = {Yao, Shunyu and Shinn, Noah and Razavi, Pedram and Narasimhan, Karthik R.},
  booktitle = {The Thirteenth International Conference on Learning Representations},
  year = {2025},
  url = {https://openreview.net/forum?id=roNSXZpUDN}
}

@article{wang2026agentnoisebench,
  title = {{AgentNoiseBench}: Benchmarking Robustness of Tool-Using {LLM} Agents Under Noisy Condition},
  author = {Wang, Ruipeng and Chen, Yuxin and Wang, Yukai and Wu, Chang and Fang, Junfeng and Cai, Xiaodong and Gu, Qi and Su, Hui and Zhang, An and Wang, Xiang and Cai, Xunliang and Chua, Tat-Seng},
  journal = {arXiv preprint arXiv:2602.11348},
  year = {2026},
  doi = {10.48550/arXiv.2602.11348},
  url = {https://arxiv.org/abs/2602.11348}
}

@inproceedings{ye2024rotbench,
  title = {{RoTBench}: A Multi-Level Benchmark for Evaluating the Robustness of Large Language Models in Tool Learning},
  author = {Ye, Junjie and Wu, Yilong and Gao, Songyang and Huang, Caishuang and Li, Sixian and Li, Guanyu and Fan, Xiaoran and Zhang, Qi and Gui, Tao and Huang, Xuanjing},
  booktitle = {Proceedings of the 2024 Conference on Empirical Methods in Natural Language Processing},
  year = {2024},
  address = {Miami, Florida, USA},
  publisher = {Association for Computational Linguistics},
  pages = {313--333},
  doi = {10.18653/v1/2024.emnlp-main.19},
  url = {https://aclanthology.org/2024.emnlp-main.19/}
}

@inproceedings{hendrycks2021mmlu,
  title = {Measuring Massive Multitask Language Understanding},
  author = {Hendrycks, Dan and Burns, Collin and Basart, Steven and Zou, Andy and Mazeika, Mantas and Song, Dawn and Steinhardt, Jacob},
  booktitle = {International Conference on Learning Representations},
  year = {2021},
  url = {https://openreview.net/forum?id=d7KBjmI3GmQ}
}

@article{srivastava2023bigbench,
  title = {Beyond the Imitation Game: Quantifying and extrapolating the capabilities of language models},
  author = {{BIG-bench authors}},
  journal = {Transactions on Machine Learning Research},
  issn = {2835-8856},
  year = {2023},
  url = {https://openreview.net/forum?id=uyTL5Bvosj}
}

@article{liang2023helm,
  title = {Holistic Evaluation of Language Models},
  author = {Liang, Percy and Bommasani, Rishi and Lee, Tony and Tsipras, Dimitris and Soylu, Dilara and Yasunaga, Michihiro and Zhang, Yian and Narayanan, Deepak and Wu, Yuhuai and Kumar, Ananya and Newman, Benjamin and Yuan, Binhang and Yan, Bobby and Zhang, Ce and Cosgrove, Christian and Manning, Christopher D. and R{\'e}, Christopher and Acosta-Navas, Diana and Hudson, Drew A. and Zelikman, Eric and Durmus, Esin and Ladhak, Faisal and Rong, Frieda and Ren, Hongyu and Yao, Huaxiu and Wang, Jue and Santhanam, Keshav and Orr, Laurel J. and Zheng, Lucia and Y{\"u}ksekg{\"o}n{\"u}l, Mert and Suzgun, Mirac and Kim, Nathan and Guha, Neel and Chatterji, Niladri S. and Khattab, Omar and Henderson, Peter and Huang, Qian and Chi, Ryan and Xie, Sang Michael and Santurkar, Shibani and Ganguli, Surya and Hashimoto, Tatsunori and Icard, Thomas and Zhang, Tianyi and Chaudhary, Vishrav and Wang, William and Li, Xuechen and Mai, Yifan and Zhang, Yuhui and Koreeda, Yuta},
  journal = {Transactions on Machine Learning Research},
  year = {2023},
  url = {https://openreview.net/forum?id=iO4LZibEqW}
}

@inproceedings{patil2024gorilla,
  title = {Gorilla: Large Language Model Connected with Massive {API}s},
  author = {Patil, Shishir G. and Zhang, Tianjun and Wang, Xin and Gonzalez, Joseph E.},
  booktitle = {Advances in Neural Information Processing Systems},
  volume = {37},
  pages = {126544--126565},
  year = {2024},
  publisher = {Curran Associates, Inc.},
  doi = {10.52202/079017-4020},
  url = {https://proceedings.neurips.cc/paper_files/paper/2024/hash/e4c61f578ff07830f5c37378dd3ecb0d-Abstract-Conference.html}
}

@article{tang2023toolalpaca,
  title = {{ToolAlpaca}: Generalized Tool Learning for Language Models with 3000 Simulated Cases},
  author = {Tang, Qiaoyu and Deng, Ziliang and Lin, Hongyu and Han, Xianpei and Liang, Qiao and Cao, Boxi and Sun, Le},
  journal = {arXiv preprint arXiv:2306.05301},
  year = {2023},
  doi = {10.48550/arXiv.2306.05301},
  url = {https://arxiv.org/abs/2306.05301}
}

@inproceedings{patil2025bfcl,
  title = {The Berkeley Function Calling Leaderboard ({BFCL}): From Tool Use to Agentic Evaluation of Large Language Models},
  author = {Patil, Shishir G and Mao, Huanzhi and Yan, Fanjia and Ji, Charlie Cheng-Jie and Suresh, Vishnu and Stoica, Ion and Gonzalez, Joseph E.},
  booktitle = {Proceedings of the 42nd International Conference on Machine Learning},
  pages = {48371--48392},
  year = {2025},
  volume = {267},
  series = {Proceedings of Machine Learning Research},
  publisher = {PMLR},
  url = {https://proceedings.mlr.press/v267/patil25a.html}
}

@inproceedings{liu2024agentbench,
  title = {{AgentBench}: Evaluating {LLM}s as Agents},
  author = {Liu, Xiao and Yu, Hao and Zhang, Hanchen and Xu, Yifan and Lei, Xuanyu and Lai, Hanyu and Gu, Yu and Ding, Hangliang and Men, Kaiwen and Yang, Kejuan and Zhang, Shudan and Deng, Xiang and Zeng, Aohan and Du, Zhengxiao and Zhang, Chenhui and Shen, Sheng and Zhang, Tianjun and Su, Yu and Sun, Huan and Huang, Minlie and Dong, Yuxiao and Tang, Jie},
  booktitle = {The Twelfth International Conference on Learning Representations},
  year = {2024},
  url = {https://openreview.net/forum?id=zAdUB0aCTQ}
}

@inproceedings{zhou2024webarena,
  title = {{WebArena}: A Realistic Web Environment for Building Autonomous Agents},
  author = {Zhou, Shuyan and Xu, Frank F. and Zhu, Hao and Zhou, Xuhui and Lo, Robert and Sridhar, Abishek and Cheng, Xianyi and Ou, Tianyue and Bisk, Yonatan and Fried, Daniel and Alon, Uri and Neubig, Graham},
  booktitle = {The Twelfth International Conference on Learning Representations},
  year = {2024},
  url = {https://openreview.net/forum?id=oKn9c6ytLx}
}

\clearpage
\appendix
\setcounter{figure}{4}

\section{Tool Environment}
\label{app:tools}

\begin{table*}[htbp]
\centering
\small
\begin{tabularx}{\linewidth}{l r Y}
\toprule
Tool group & \# tools & Tool IDs \\
\midrule
\texttt{calendar\_ops} & 4 & \texttt{cancel\_calendar\_event}, \texttt{check\_calendar\_availability}, \texttt{create\_calendar\_event}, \texttt{update\_calendar\_event} \\
\texttt{commerce\_ops} & 4 & \texttt{compare\_product\_price}, \texttt{estimate\_shipping}, \texttt{lookup\_product}, \texttt{track\_order} \\
\texttt{conversion} & 2 & \texttt{convert\_currency}, \texttt{convert\_unit} \\
\texttt{crm\_ops} & 4 & \texttt{create\_support\_ticket}, \texttt{lookup\_customer}, \texttt{lookup\_invoice}, \texttt{update\_ticket\_status} \\
\texttt{database\_ops} & 4 & \texttt{aggregate\_table}, \texttt{lookup\_record}, \texttt{query\_table}, \texttt{validate\_record} \\
\texttt{email\_ops} & 4 & \texttt{classify\_email}, \texttt{draft\_email}, \texttt{search\_email}, \texttt{summarize\_email} \\
\texttt{file\_ops} & 4 & \texttt{convert\_file\_format}, \texttt{extract\_file\_metadata}, \texttt{search\_file}, \texttt{summarize\_file} \\
\texttt{list\_ops} & 2 & \texttt{filter\_items}, \texttt{sort\_items} \\
\texttt{route\_travel} & 2 & \texttt{calculate\_route}, \texttt{estimate\_travel\_time} \\
\texttt{search\_retrieval} & 3 & \texttt{search\_document}, \texttt{search\_faq}, \texttt{search\_policy} \\
\texttt{security\_ops} & 4 & \texttt{audit\_login\_events}, \texttt{check\_permission}, \texttt{revoke\_session}, \texttt{rotate\_api\_key} \\
\texttt{weather\_env} & 3 & \texttt{get\_air\_quality}, \texttt{get\_temperature}, \texttt{get\_weather} \\
\bottomrule
\end{tabularx}
\caption{Complete deterministic local-tool environment in \ToolRobustBench. Each listed function has a structured schema, deterministic executor, and expected output format; the main single-family experiments sample 16 of these 40 tools with fixed seed 20260511.}
\label{tab:tool-env}
\end{table*}

\section{Data Construction Details}
\label{app:data-construction}

\subsection{Clean Seed Anchor}

Table~\ref{tab:clean-seed-example} presents an anchor taken from the deterministic task set. The request distinguishes a route computation from its neighboring travel-time function; the stored arguments are executed before perturbation to obtain the expected result. Perturbed variants derived from this anchor preserve these targets unless an observation or runtime template supplies a distinct post-execution judgment target.

\begin{table*}[htbp]
\centering
\small
\begin{tabularx}{\linewidth}{l Y}
\toprule
Field & Clean seed value \\
\midrule
User request & \texttt{I need to get from hotel to expo using metro; give me the way to go, not just the time.} \\
Candidate schema summary & Relevant sampled siblings include \texttt{calculate\_route(origin, target, profile)} and \texttt{estimate\_travel\_time(origin, target, profile)}. \\
Gold tool & \texttt{calculate\_route} \\
Gold arguments & \texttt{\{"origin": "hotel", "target": "expo", "profile": "metro"\}} \\
Deterministic output & \texttt{\{"origin": "hotel", "target": "expo", "profile": "metro", "distance\_km": 9, "duration\_min": 22\}} \\
Expected result & The returned route record for a metro trip from \texttt{hotel} to \texttt{expo}, including distance 9 km and duration 22 min. \\
\bottomrule
\end{tabularx}
\caption{A clean seed task used as a perturbation anchor. Values are taken from the deterministic generated task record \texttt{route\_01}.}
\label{tab:clean-seed-example}
\end{table*}

\subsection{Subtype Template Operations}

Table~\ref{tab:subtype-details} expands the main-text specification with the input field altered by each subtype and representative deterministic operations implemented by the template catalog. The expected failure modes are diagnostic hypotheses used for attribution; they do not require that every failed model output take that form.

\begin{table*}[htbp]
\centering
\scriptsize
\begin{tabularx}{\linewidth}{l l Y Y}
\toprule
ID & Modified field & Template operation in the construction catalog & Intended stage and expected failure mode \\
\midrule
A1 & Exposed tool name & Substitute a readable alias, semi-opaque identifier, or internal-looking name while retaining the canonical mapping. & S1; incorrect tool selection under surface shift. \\
A2 & Tool description & Rewrite or abstract descriptions to remove directly matching phrases without changing the coarse function. & S2; wrong tool after weak schema grounding. \\
A3 & Names/descriptions of siblings & Add sibling similarity cues or near-colliding family names/descriptions. & S1/S2; confusion between neighboring tools. \\
B1 & User request wording & Paraphrase, indirect style transfer, reposition constraints, or scatter implied cues across sentences. & S2/S3; lost intended tool or argument cue. \\
B2 & User request slots & Make quantities vague, omit modifiers, or render slots implicit while retaining a designated recovery path. & S3; missing or incorrect argument. \\
B3 & User request context & Add redundant context, a semantically nearby distractor, or a secondary goal. & S1/S3; redirected tool or arguments. \\
B4 & User request constraints & Introduce a contradictory constraint or late revision with a designated primary target. & S1/S3; unresolved intent conflict. \\
C1 & Returned observation structure & Rename keys, reorder fields, or shift nesting of an executed return packet. & S4; invalid-format or incorrect-result recovery. \\
C2 & Returned evidence fields & Drop metadata, sparsify results, omit a high-value field, or truncate supporting evidence. & S4; incorrect evidence recovery from incomplete returned information. \\
C3 & Returned values & Insert stale values, conflicting supporting facts, unit mismatches, or an incorrect fact field. & S4; acceptance of corrupted evidence. \\
C4 & Returned candidate evidence & Add redundant wrappers, alternative candidates, auxiliary evidence, or misleading notes. & S4; selection of a competing result. \\
D1 & Execution availability state & Supply fallback evidence, service-unavailable state, or timeout-after-partial-stream feedback. & S4; incorrect status or retry judgment. \\
D2 & Runtime attempt history & Add jitter metadata, stale-versus-fresh fragments, or conflicting attempts. & S4; incorrect final-state resolution. \\
D3 & Failure feedback packet & Provide vague retry cues, unreliable failure labels, or hidden partial evidence with diagnostic decoys. & S4; incorrect failure-type, retryability, or retained-result decision. \\
\bottomrule
\end{tabularx}
\caption{Template-level construction details for all A1--D3 subtypes. Subtype names match the implementation catalog; operations are deterministic transformations of an anchored instance or its returned/runtime packet.}
\label{tab:subtype-details}
\end{table*}

\subsection{Severity Criteria}

\begin{table*}[htbp]
\centering
\small
\begin{tabularx}{\linewidth}{l Y Y}
\toprule
Severity & General construction criterion & Diagnostic interpretation \\
\midrule
Light & Introduce one weak or surface-level change, such as a readable alias, minor rephrasing, small structure change, or transparent runtime cue. & Target remains highly recoverable; tests sensitivity to small presentation shifts. \\
Medium & Increase ambiguity, remove or compete with salient evidence, or expose an uncertain execution state. & Requires robust grounding or recovery rather than surface matching. \\
Heavy & Apply the strongest controlled subtype operation, possibly with an amplifier operation, while retaining a defined target or runtime decision. & Tests behavior under substantial stress; it is not an assertion of monotonically lower model accuracy. \\
\bottomrule
\end{tabularx}
\caption{General severity criteria for template expansion. The generation pipeline validates monotonic static strength through a preflight gate before evaluation; severity controls perturbation strength at construction time rather than imposing an empirical accuracy ordering.}
\label{tab:severity-criteria}
\end{table*}

\subsection{Gold and Diagnostic Fields}

\begin{table*}[htbp]
\centering
\scriptsize
\begin{tabularx}{\linewidth}{l l Y}
\toprule
Paper field & Stored record field & Meaning in construction and scoring \\
\midrule
\texttt{gold\_tool} & \texttt{gold\_canonical\_tool\_id} & Canonical tool identity fixed by the clean seed anchor. \\
\texttt{gold\_arguments} & \texttt{expected\_arguments} & Structured argument object executed on the clean tool. \\
\texttt{expected\_result} & \texttt{expected\_result} & Deterministic clean result or target content to recover. \\
\texttt{observed\_error} & \texttt{observed\_failure\_tag} & Failure symptom observed at the evaluated output. \\
\texttt{primary\_error} & \texttt{primary\_failure\_tag} & Principal failure source assigned after cascade backtracking. \\
\texttt{earliest\_failed\_stage} & \texttt{first\_failure\_stage} & Earliest S1--S4 diagnostic stage at which behavior fails. \\
\texttt{error\_source\_stage} & \texttt{cascade\_from\_stage} & Source stage when a downstream symptom is attributed upstream. \\
\texttt{cascade indicator} & \texttt{is\_cascade\_failure} & Whether observed and source-attributed failure differ. \\
\texttt{boundary-violation indicator} & \texttt{boundary\_violation\_flag} & Whether the attributed source falls outside the intended family boundary. \\
\texttt{runtime status} & \texttt{decision\_status}, \texttt{expected\_statuses} & Predicted and accepted runtime states for post-execution instances. \\
\texttt{retryability} & \texttt{decision\_retryable}, \texttt{expected\_retryable} & Predicted and expected retry decision when runtime feedback applies. \\
\texttt{failure type} & \texttt{decision\_failure\_type}, \texttt{expected\_failure\_types} & Predicted and accepted runtime-failure category when applicable. \\
\bottomrule
\end{tabularx}
\caption{Construction and diagnostic label fields preserved in raw evaluation records. Runtime-decision fields apply to the post-execution portion of the benchmark.}
\label{tab:construction-labels}
\end{table*}

\subsection{Deterministic Generation Artifacts}

Construction is implemented with deterministic templates rather than generated evaluation instances from an external language model. GPT-assisted drafting is limited to natural-language template diversification before inclusion; after a template is accepted, clean anchors, perturbations, executor outputs, and diagnostic labels are generated by deterministic code. Implementation files are under \path{src/toolrobust/}: dataset builders create clean anchors and executor results, tool definitions define the registry, and perturbation modules specify subtype operators, operator strength scores, severity-specific transformations, and the preflight severity gate. Serialized construction artifacts are stored under \path{data/generated/}, including the preflight severity-gate report; raw model-output records used for the main count appear under \path{data/results/}.

\section{Evaluation Validity Details}
\label{app:validity}

We perform an integrated validation audit on 140 stratified single-family records, selecting ten records from each subtype A1--D3. The audit contains 140 records in total, but agreement is label-specific. We exclude \texttt{unclear} judgments from each label-level agreement computation and apply the prespecified \texttt{not\_applicable} convention to non-runtime decision fields; therefore, the valid comparison count differs across labels. The completed human annotation covers data-construction validity, task success, error attribution, boundary judgments, and runtime-decision labels. The human annotators are professionally trained annotators with more than two years of experience in data annotation and evaluation tasks. A blind LLM annotator, GPT-4.5 accessed through API, is evaluated on the same visible record evidence as a secondary consistency check rather than ground truth; runtime-only fields are normalized to \texttt{not\_applicable} for A/B/C records before computing agreement. The audit yields human--scorer agreement of 0.895 ($\kappa=0.777$) for task success, 0.814 ($\kappa=0.755$) for primary error, 0.907 ($\kappa=0.734$) for status, 0.957 ($\kappa=0.873$) for retryability, and 0.921 ($\kappa=0.772$) for failure type. These values provide an external check on the scorer used for the full evaluation.

\begin{table*}[htbp]
\centering
\small
\begin{tabular}{lrrr}
\toprule
Validation comparison & Label & Valid $n$ & Agreement / $\kappa$ \\
\midrule
Human--scorer & Task success & 86 & 0.895 / 0.777 \\
Human--scorer & Observed error & 86 & 0.837 / 0.790 \\
Human--scorer & Primary error & 86 & 0.814 / 0.755 \\
Human--scorer & Status & 140 & 0.907 / 0.734 \\
Human--scorer & Retryability & 140 & 0.957 / 0.873 \\
Human--scorer & Failure type & 139 & 0.921 / 0.772 \\
\midrule
Human--LLM & Task success & 86 & 0.802 / 0.565 \\
Human--LLM & Status & 140 & 0.900 / 0.725 \\
Human--LLM & Retryability & 140 & 0.971 / 0.917 \\
Human--LLM & Failure type & 139 & 0.892 / 0.707 \\
\bottomrule
\end{tabular}
\caption{Integrated validation on 140 stratified human-audited records. Valid $n$ denotes the number of comparable records for each label after excluding \texttt{unclear} judgments and applying the prespecified \texttt{not\_applicable} convention for non-runtime decision fields.}
\label{tab:integrated-validation}
\end{table*}

The principal findings persist after scorer normalization and protocol-compatibility adjustments. For runtime-environment perturbations, failure-type alias normalization prevents semantically equivalent labels from being scored as incorrect solely because their surface forms differ. After normalization, family-D results remain differentiated: \texttt{deepseek-v4-pro} reaches 0.748, \texttt{claude-sonnet-4-6} reaches 0.745, \texttt{qwen-plus} reaches 0.713, and \texttt{gpt-5.4-mini} reaches 0.708. Normalization thus addresses a scoring artifact without erasing observed model differences.

For post-execution decisions in the tool-output/observation and runtime-environment families, compatibility checks test whether empty or invalid structured outputs result from backend token-budget or protocol failures. Raw records remain the traceable basis of the primary results; any compatibility-adjusted estimates are reported separately as sensitivity analyses. Appendix Figures~\ref{fig:app-alias}, \ref{fig:app-nonmonotonic}, and \ref{fig:app-gemini-profile} provide perturbation-calibration and scorer-ablation diagnostics.

\section{Evidence Table}
\label{app:evidence}

\begin{table*}[htbp]
\centering
\small
\begin{adjustbox}{width=\linewidth}
\begin{tabular}{p{0.28\linewidth} p{0.20\linewidth} p{0.25\linewidth} p{0.07\linewidth} p{0.12\linewidth}}
\toprule
Data source & Citable metrics & Corresponding conclusion & Main text & Figure \\
\midrule
All-family raw records & sample count, success, failure tag, stage fields & Complete raw records for seven models, totaling 15,456 instances & Yes & Figs.~\ref{fig:family-heatmap}--\ref{fig:failure-taxonomy}; Fig.~\ref{fig:app-cascade-boundary} \\
Overall summaries & condition-level summary & Auxiliary sanity checks; main-text values are recomputed from raw data & No & No \\
Family summaries & family/severity success & Agreement with raw records; main-text values are recomputed from raw data & Yes & Figure~\ref{fig:family-heatmap}, Appendix Figure~\ref{fig:app-severity} \\
Subtype summaries & subtype success & Difficulty of C2, D2, and B3 and non-monotonic response analysis & Yes & Appendix Figures~\ref{fig:app-subtype}, \ref{fig:app-nonmonotonic}, \ref{fig:app-gemini-profile} \\
Failure-taxonomy records & primary failure tag & Concentration of family-specific primary failure modes & Yes & Figure~\ref{fig:failure-taxonomy} \\
Boundary reports & cascade rate, boundary-violation rate & Distinction between cascades and boundary violations & Yes & Appendix Figure~\ref{fig:app-cascade-boundary} \\
C2 recoverability outputs & model-proxy solvability, human YES/PARTIAL/NO labels & Evidence-loss perturbations remain recoverable for many instances & Yes & Table~\ref{tab:c2-recoverability} \\
Counterfactual attribution outputs & tier-level and family-level recovery rates & Primary-error labels identify causal failure sources in most cases & Yes & Table~\ref{tab:counterfactual-attribution} \\
Targeted intervention outputs & B-family S3-CoT and A-family S1-alias interventions & Diagnostic labels reject mismatched fixes and prescribe useful repairs & Yes & Table~\ref{tab:diagnostic-interventions} \\
Mixed-family raw records & mixed-family success, failure tag, cascade fields & Medium/heavy interactions for five representative pairwise combinations & Yes & Figure~\ref{fig:cross-vs-single}, Appendix Figures~\ref{fig:app-cross-success}--\ref{fig:app-cross-failure} \\
Mixed-vs-single summary & mixed-family success, weaker constituent, delta & Comparison between mixed-family and constituent-family performance & Yes & Figure~\ref{fig:cross-vs-single} \\
Mixed-family plot manifest & retained plot type, source plot directory & Retained plot types for each model $\times$ combination & No & Appendix Figures~\ref{fig:app-cross-model-sheet}--\ref{fig:app-gpt-cross-matrix} \\
Evaluation-validity notes & scorer correction and compatibility diagnosis & Validity checks and ablation design & Yes & Appendix Figure~\ref{fig:app-alias} \\
Integrated human audit & 140 stratified annotation records, LLM cross-check, agreement tables & Human--scorer and human--LLM validation of scorer-facing labels & Yes & Table~\ref{tab:integrated-validation}, Appendix~\ref{app:validity} \\
Run manifests & sampled tools, dataset scope, family config & 16 sampled tools, complete environment of 40 tools/12 groups, single-step setting & Yes & Appendix Table~\ref{tab:tool-env} \\
Figure assets & main figures & Four core main-text figures, with secondary diagnostics & Yes & Figs.~\ref{fig:pipeline}, \ref{fig:family-heatmap}, \ref{fig:failure-taxonomy}, \ref{fig:cross-vs-single}; Figs.~\ref{fig:app-stage2}--\ref{fig:app-gpt-cross-matrix} \\
\bottomrule
\end{tabular}
\end{adjustbox}
\caption{Evidence table linking data sources to reported metrics, claims, and figures.}
\label{tab:evidence}
\end{table*}

\section{Complete Model Result Tables}
\label{app:model-results}

Table~\ref{tab:family-overall} summarizes each model's perturbed macro-average across the four families.

\begin{table*}[htbp]
\centering
\small
\begin{tabular}{lrrrrr}
\toprule
Model & Tool-interface & User-intent & Tool-output/observation & Runtime-environment & Overall \\
\midrule
\texttt{claude-sonnet-4-6} & 0.968 & 0.785 & 0.568 & 0.745 & 0.766 \\
\texttt{deepseek-v4-pro} & 0.926 & 0.809 & 0.512 & 0.748 & 0.749 \\
\texttt{gpt-5.4-mini} & 0.856 & 0.781 & 0.503 & 0.708 & 0.712 \\
\texttt{qwen-plus} & 0.931 & 0.802 & 0.380 & 0.713 & 0.706 \\
\texttt{gemini-2.5-pro} & 0.900 & 0.748 & 0.481 & 0.637 & 0.692 \\
\texttt{deepseek-v4-flash} & 0.958 & 0.783 & 0.319 & 0.616 & 0.669 \\
\texttt{gemini-2.5-flash} & 0.889 & 0.700 & 0.420 & 0.648 & 0.664 \\
\bottomrule
\end{tabular}
\caption{Family-level robustness for each model.}
\label{tab:family-overall}
\end{table*}

Tables~\ref{tab:claude-severity}--\ref{tab:qwen-severity} report success rates for each family $\times$ severity condition. All values are recomputed directly from raw \texttt{records.csv}.

\begin{table}[htbp]
\centering
\small
\begin{tabular}{lrrrr}
\toprule
Family & Clean & Light & Medium & Heavy \\
\midrule
Interface & 1.000 & 1.000 & 0.972 & 0.931 \\
Intent & 1.000 & 0.974 & 0.870 & 0.510 \\
Output & 1.000 & 0.661 & 0.625 & 0.417 \\
Runtime & 1.000 & 0.972 & 0.708 & 0.556 \\
\bottomrule
\end{tabular}
\caption{Family $\times$ severity success rates for \texttt{claude-sonnet-4-6}.}
\label{tab:claude-severity}
\end{table}

\begin{table}[htbp]
\centering
\small
\begin{tabular}{lrrrr}
\toprule
Family & Clean & Light & Medium & Heavy \\
\midrule
Interface & 1.000 & 1.000 & 0.972 & 0.903 \\
Intent & 0.979 & 0.953 & 0.839 & 0.557 \\
Output & 0.750 & 0.406 & 0.370 & 0.182 \\
Runtime & 0.979 & 1.000 & 0.521 & 0.326 \\
\bottomrule
\end{tabular}
\caption{Family $\times$ severity success rates for \texttt{deepseek-v4-flash}.}
\label{tab:deepseek-flash-severity}
\end{table}

\begin{table}[htbp]
\centering
\small
\begin{tabular}{lrrrr}
\toprule
Family & Clean & Light & Medium & Heavy \\
\midrule
Interface & 1.000 & 1.000 & 0.924 & 0.854 \\
Intent & 1.000 & 0.990 & 0.885 & 0.552 \\
Output & 1.000 & 0.719 & 0.589 & 0.229 \\
Runtime & 1.000 & 1.000 & 0.729 & 0.514 \\
\bottomrule
\end{tabular}
\caption{Family $\times$ severity success rates for \texttt{deepseek-v4-pro}.}
\label{tab:deepseek-pro-severity}
\end{table}

\begin{table}[htbp]
\centering
\small
\begin{tabular}{lrrrr}
\toprule
Family & Clean & Light & Medium & Heavy \\
\midrule
Interface & 0.958 & 0.986 & 0.875 & 0.806 \\
Intent & 0.958 & 0.901 & 0.823 & 0.375 \\
Output & 0.979 & 0.635 & 0.495 & 0.130 \\
Runtime & 0.958 & 0.958 & 0.431 & 0.556 \\
\bottomrule
\end{tabular}
\caption{Family $\times$ severity success rates for \texttt{gemini-2.5-flash}.}
\label{tab:gemini-flash-severity}
\end{table}

\begin{table}[htbp]
\centering
\small
\begin{tabular}{lrrrr}
\toprule
Family & Clean & Light & Medium & Heavy \\
\midrule
Interface & 1.000 & 0.979 & 0.903 & 0.819 \\
Intent & 1.000 & 0.932 & 0.875 & 0.438 \\
Output & 0.979 & 0.693 & 0.562 & 0.188 \\
Runtime & 0.938 & 0.958 & 0.486 & 0.465 \\
\bottomrule
\end{tabular}
\caption{Family $\times$ severity success rates for \texttt{gemini-2.5-pro}.}
\label{tab:gemini-pro-severity}
\end{table}

\begin{table}[htbp]
\centering
\small
\begin{tabular}{lrrrr}
\toprule
Family & Clean & Light & Medium & Heavy \\
\midrule
Interface & 1.000 & 1.000 & 0.840 & 0.729 \\
Intent & 0.979 & 0.969 & 0.875 & 0.500 \\
Output & 0.979 & 0.703 & 0.641 & 0.167 \\
Runtime & 0.979 & 0.938 & 0.618 & 0.569 \\
\bottomrule
\end{tabular}
\caption{Family $\times$ severity success rates for \texttt{gpt-5.4-mini}.}
\label{tab:gpt-mini-severity}
\end{table}

\begin{table}[htbp]
\centering
\small
\begin{tabular}{lrrrr}
\toprule
Family & Clean & Light & Medium & Heavy \\
\midrule
Interface & 1.000 & 1.000 & 0.944 & 0.847 \\
Intent & 1.000 & 0.984 & 0.896 & 0.526 \\
Output & 1.000 & 0.526 & 0.417 & 0.198 \\
Runtime & 1.000 & 0.854 & 0.729 & 0.556 \\
\bottomrule
\end{tabular}
\caption{Family $\times$ severity success rates for \texttt{qwen-plus}.}
\label{tab:qwen-severity}
\end{table}

\begin{table*}[htbp]
\centering
\small
\begin{tabular}{lrrrrr}
\toprule
Family & Failures & Cascade rate & Expected-primary match & Spillover match & Boundary violation \\
\midrule
Tool-interface & 247 & 0.352 & 0.891 & 0.352 & 0.109 \\
User-intent & 917 & 0.108 & 0.957 & 0.108 & 0.043 \\
Tool-output/observation & 2198 & 0.003 & 0.931 & 0.000 & 0.069 \\
Runtime-environment & 944 & 0.638 & 0.994 & 0.638 & 0.006 \\
\bottomrule
\end{tabular}
\caption{Family-level attribution statistics over failed instances. Cascade rate measures observed/primary disagreement after backtracking; boundary violation measures attributed sources outside the intended family boundary.}
\label{tab:boundary-stats}
\end{table*}

\FloatBarrier
\section{Additional Diagnostic Figures}
\label{app:figures}

The appendix figures provide additional diagnostics for dataset coverage, clean-versus-robustness comparisons, severity effects, subtype difficulty, post-execution behavior, scorer normalization, cascade/boundary behavior, and mixed-family interactions. Figures~\ref{fig:app-leaderboard} and \ref{fig:app-cascade-boundary} retain the leaderboard and cascade-boundary diagnostics.

Figures~\ref{fig:app-dataset-coverage}--\ref{fig:app-gemini-profile} report dataset coverage, clean-versus-robustness comparisons, severity curves, subtype difficulty, post-execution diagnostics, alias ablation, non-monotonic cases, and the Gemini-Pro execution profile. Figures~\ref{fig:app-cross-success}--\ref{fig:app-gpt-cross-matrix} report mixed-family success and attribution results.

Figures~\ref{fig:app-interface-sheet}--\ref{fig:app-gpt-cross-matrix} provide diagnostic sheets and mixed-family summaries, including the full composition space available for \texttt{gpt-5.4-mini}.

\ifincludeappendixfigures
\begin{figure*}[t]
\centering
\includegraphics[width=0.95\linewidth]{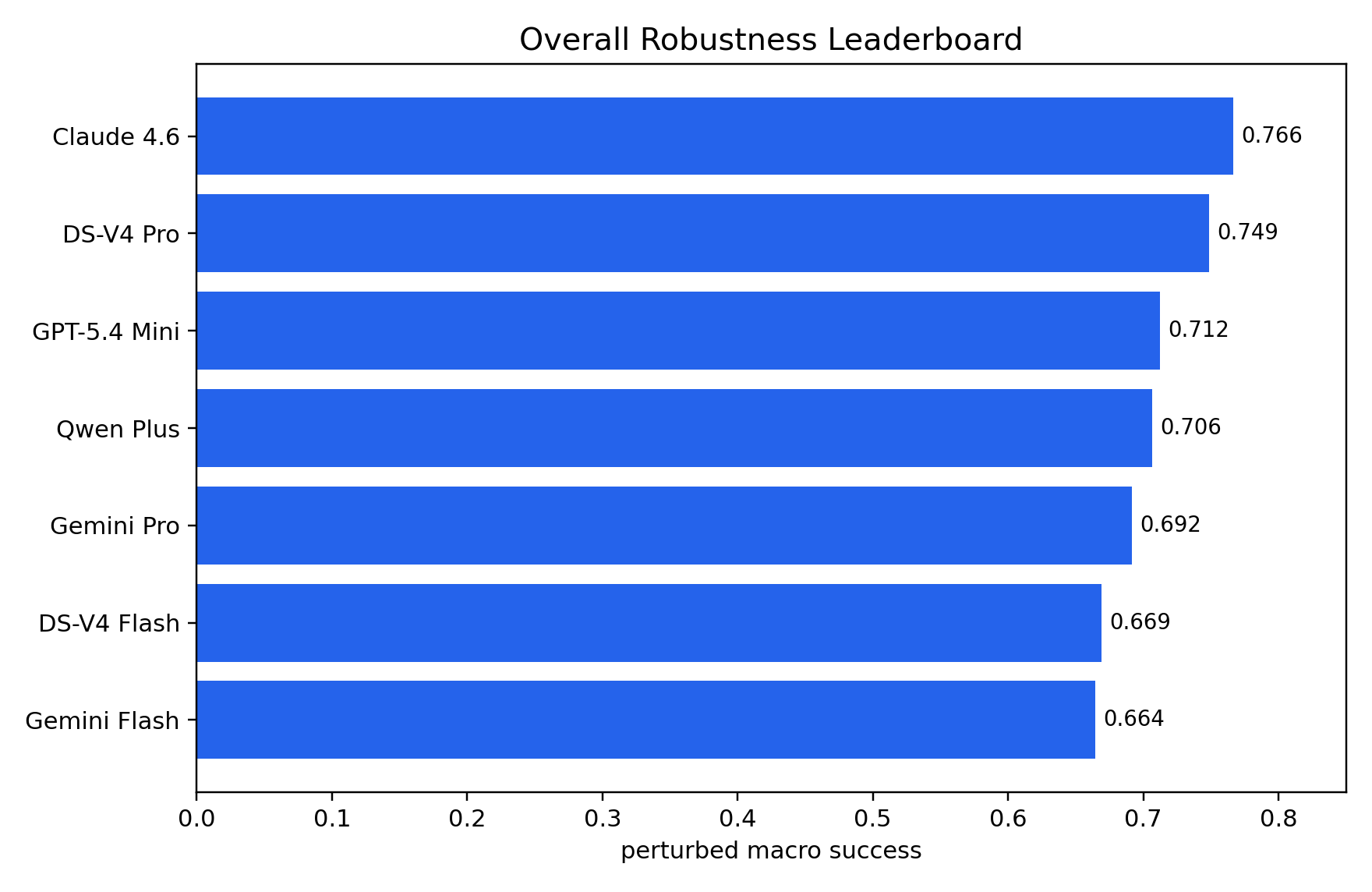}
\caption{Overall robustness leaderboard with clean success, macro-averaged perturbed performance, and the four family-level robustness components.}
\label{fig:app-leaderboard}
\end{figure*}

\begin{figure*}[t]
\centering
\includegraphics[width=0.95\linewidth]{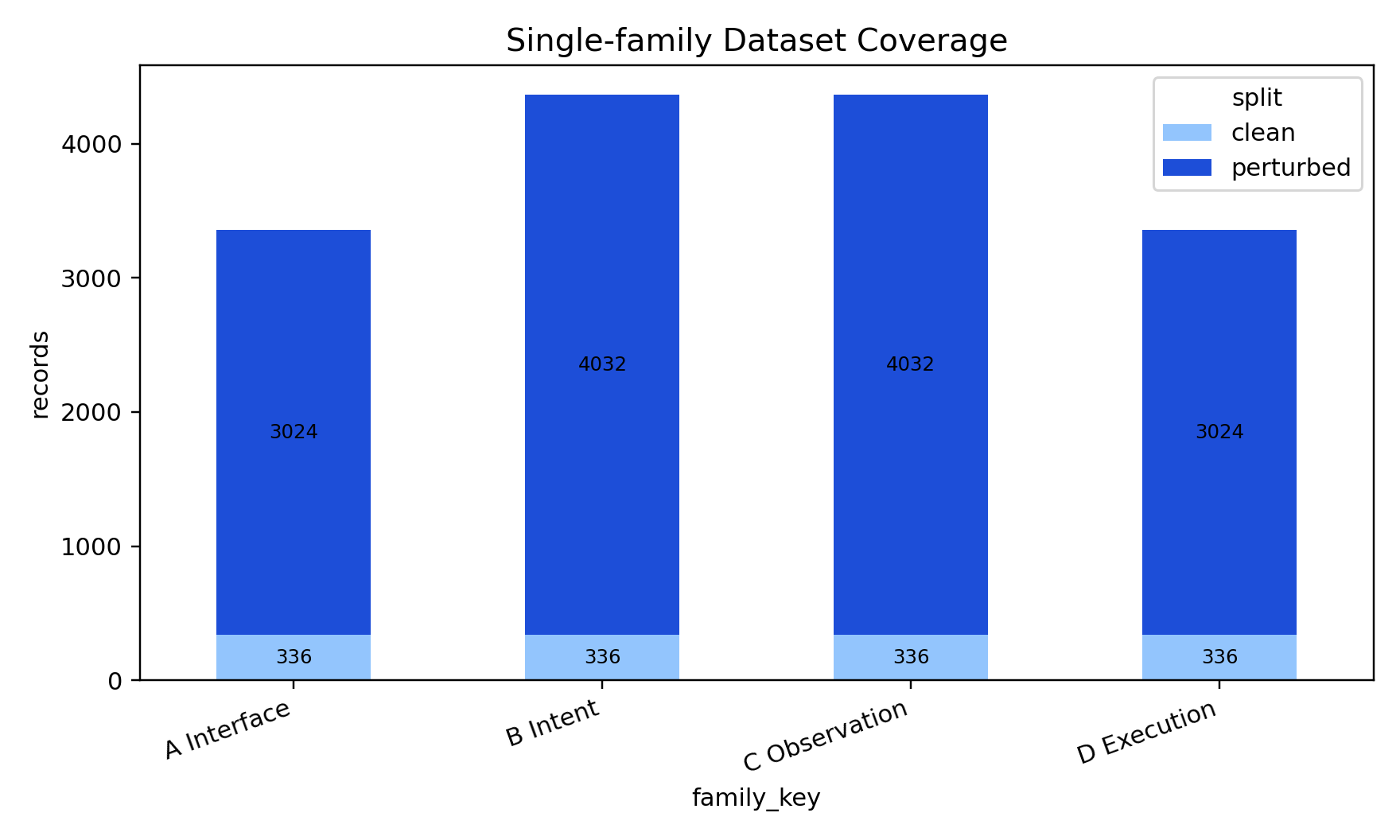}
\caption{Coverage of the deterministic evaluation data over models, sampled tools, perturbation families, subtypes, and severity conditions.}
\label{fig:app-dataset-coverage}
\end{figure*}

\begin{figure*}[t]
\centering
\includegraphics[width=0.95\linewidth]{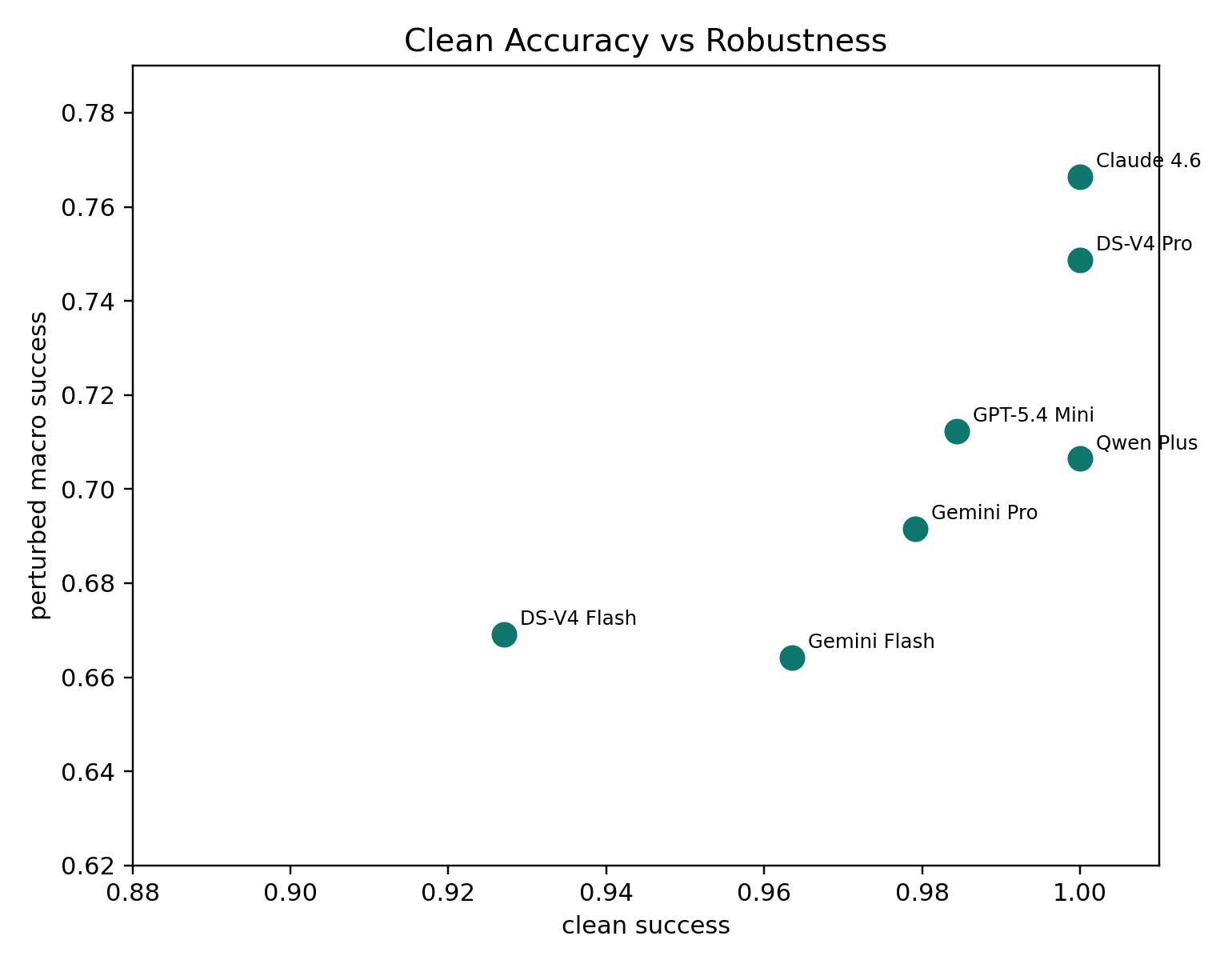}
\caption{Clean-setting E2E success compared with overall perturbed robustness; the gap indicates that clean accuracy does not predict stage-wise robustness.}
\label{fig:app-clean-robustness}
\end{figure*}

\begin{figure*}[t]
\centering
\includegraphics[width=0.95\linewidth]{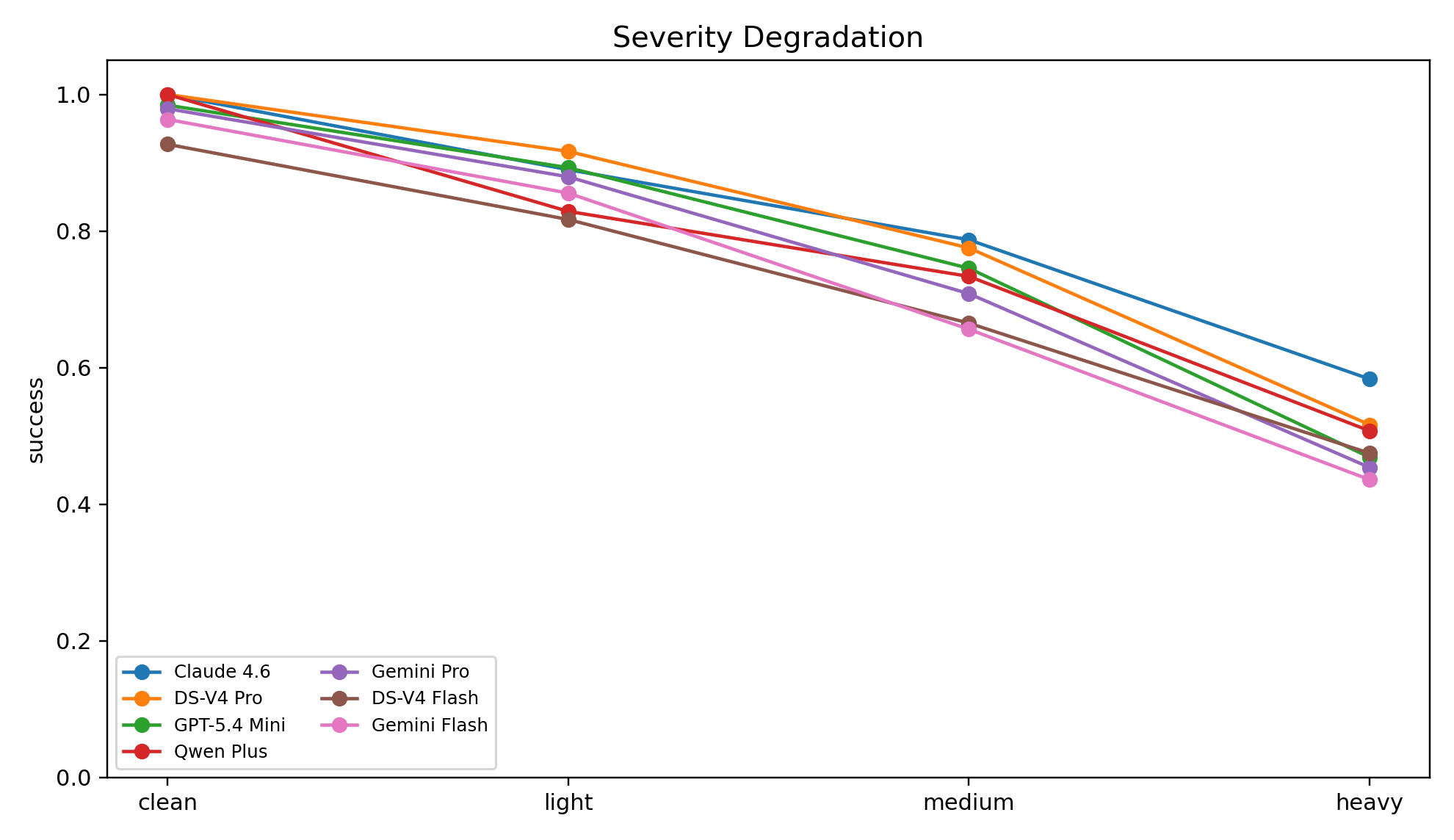}
\caption{Success rates across clean, light, medium, and heavy conditions; severity is a calibrated input-side stress level and need not force monotonic model accuracy for every subtype.}
\label{fig:app-severity}
\end{figure*}

\begin{figure*}[t]
\centering
\includegraphics[width=0.95\linewidth]{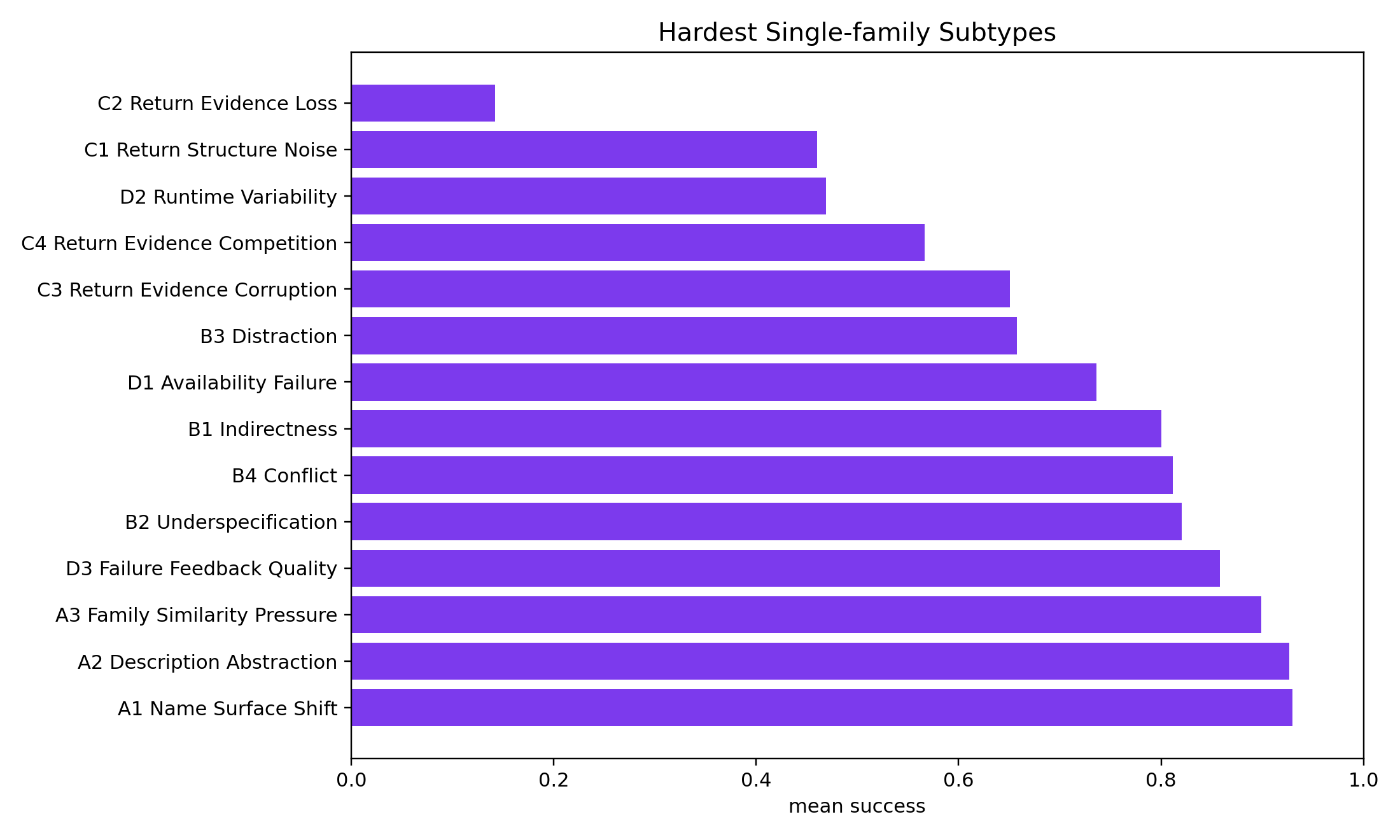}
\caption{Robustness by perturbation subtype A1--D3; family-C evidence loss and competition expose the most difficult observation-recovery behavior.}
\label{fig:app-subtype}
\end{figure*}

\begin{figure*}[t]
\centering
\includegraphics[width=0.95\linewidth]{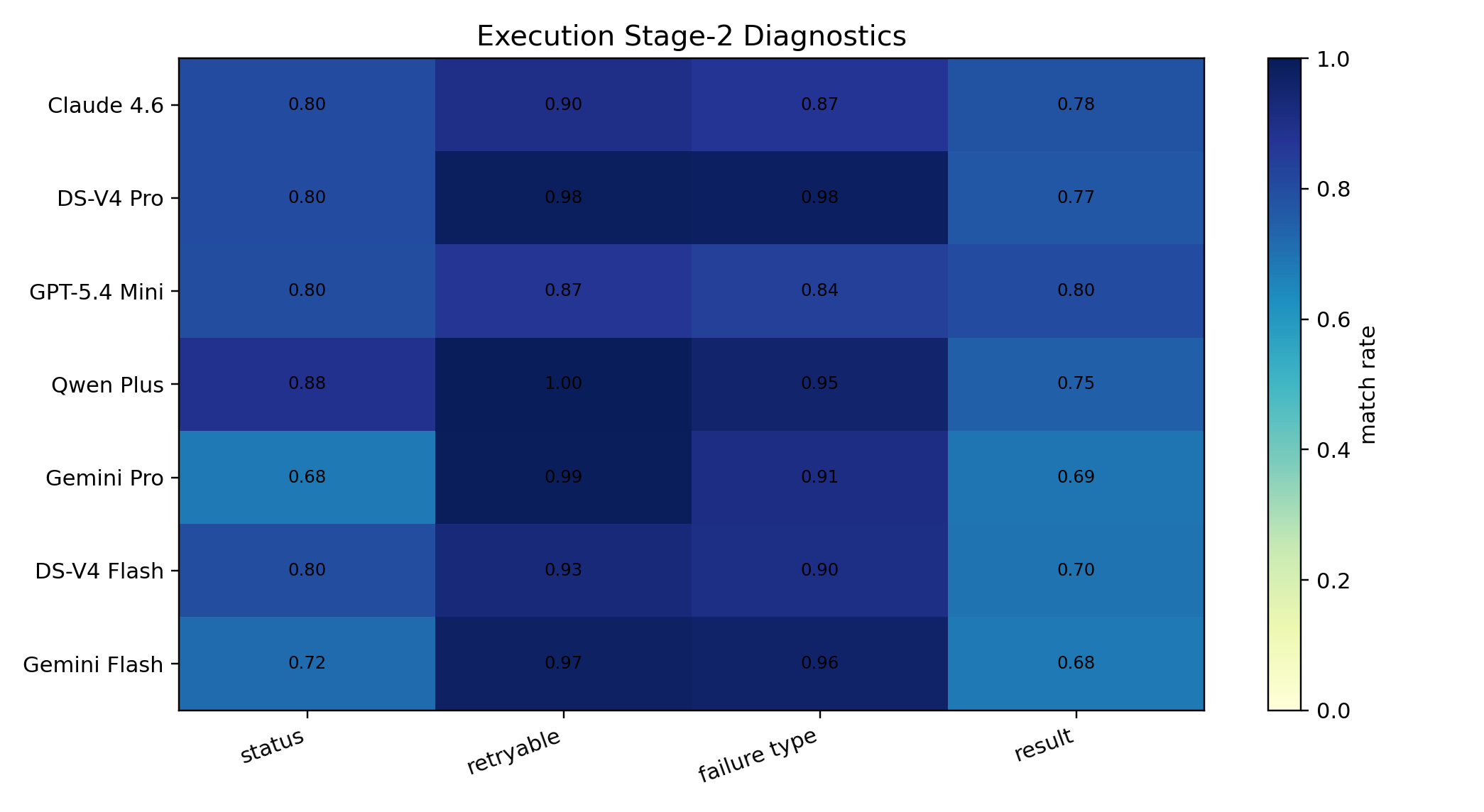}
\caption{Post-execution decision accuracy for status, retryability, failure type, and final result in tool-output/observation and runtime-environment evaluation.}
\label{fig:app-stage2}
\end{figure*}

\begin{figure*}[t]
\centering
\includegraphics[width=0.95\linewidth]{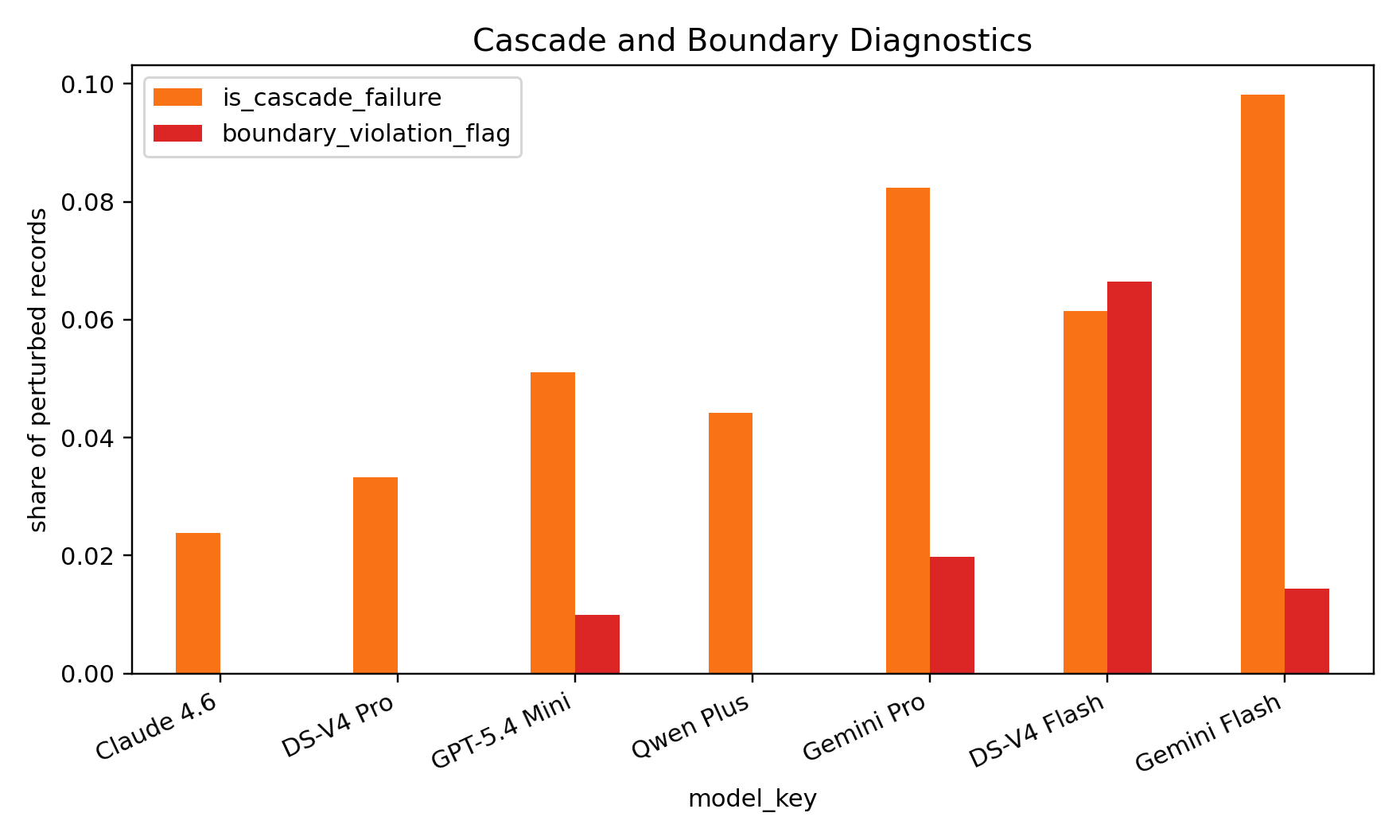}
\caption{Cascade and boundary diagnostics by family. Cascade rate captures downstream symptoms whose primary source backtracks upstream; boundary-violation rate captures attributed sources beyond the intended family target.}
\label{fig:app-cascade-boundary}
\end{figure*}

\begin{figure*}[t]
\centering
\includegraphics[width=0.95\linewidth]{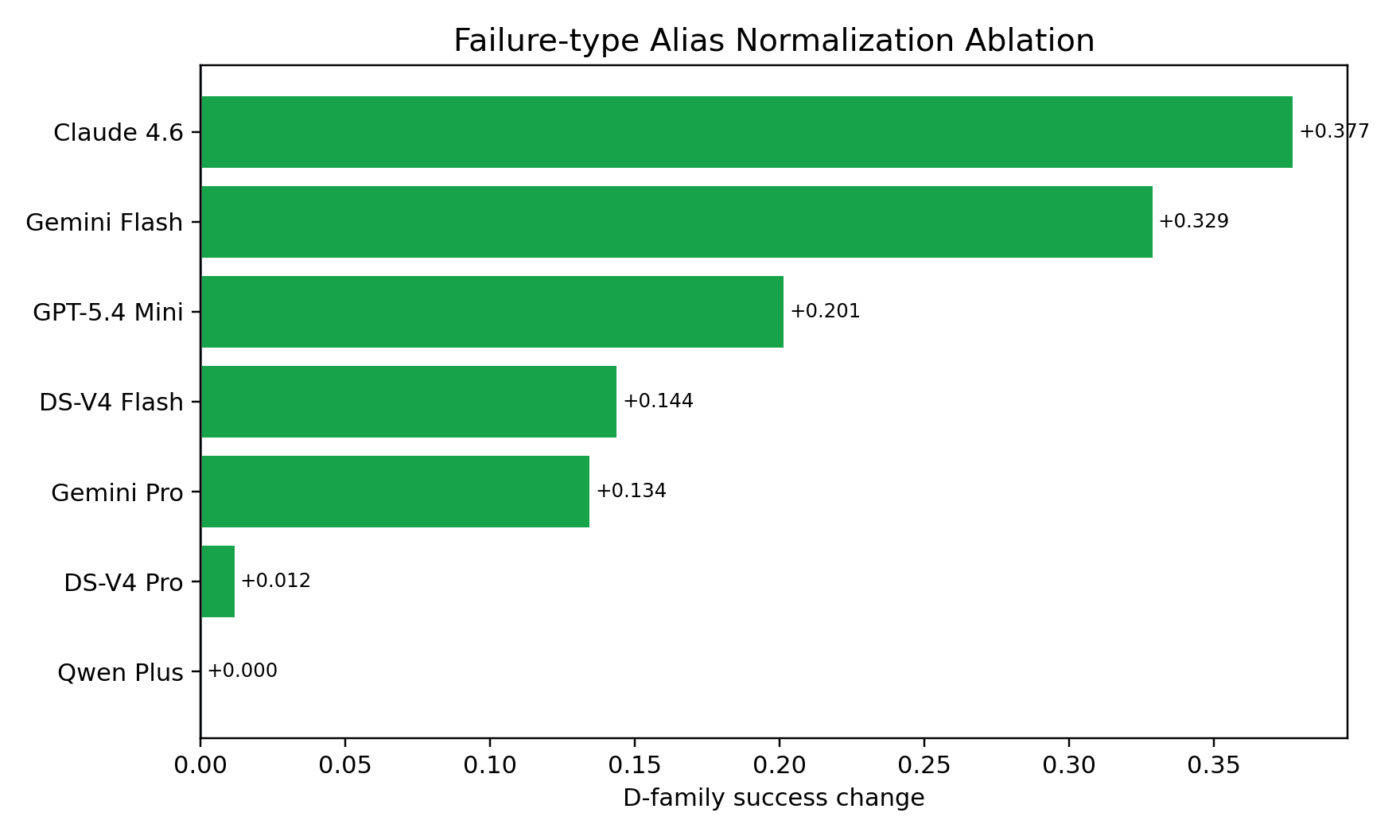}
\caption{Failure-type alias normalization ablation for runtime decisions, comparing equivalent-label handling before and after model-independent normalization.}
\label{fig:app-alias}
\end{figure*}

\begin{figure*}[t]
\centering
\includegraphics[width=0.95\linewidth]{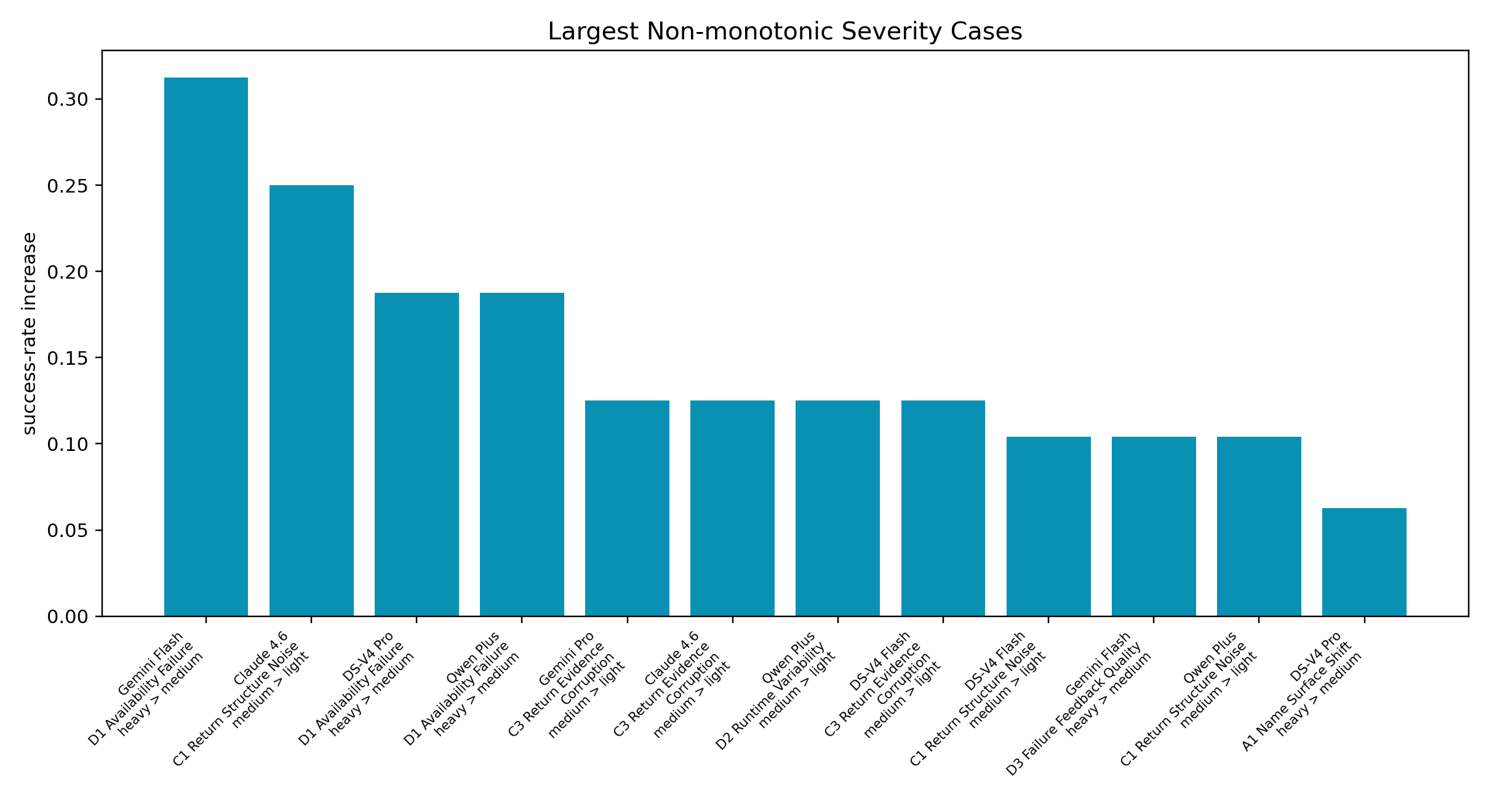}
\caption{Selected non-monotonic severity-response cases, in which more explicit failure or contamination cues can improve classification despite greater input-side stress.}
\label{fig:app-nonmonotonic}
\end{figure*}

\begin{figure*}[t]
\centering
\includegraphics[width=0.95\linewidth]{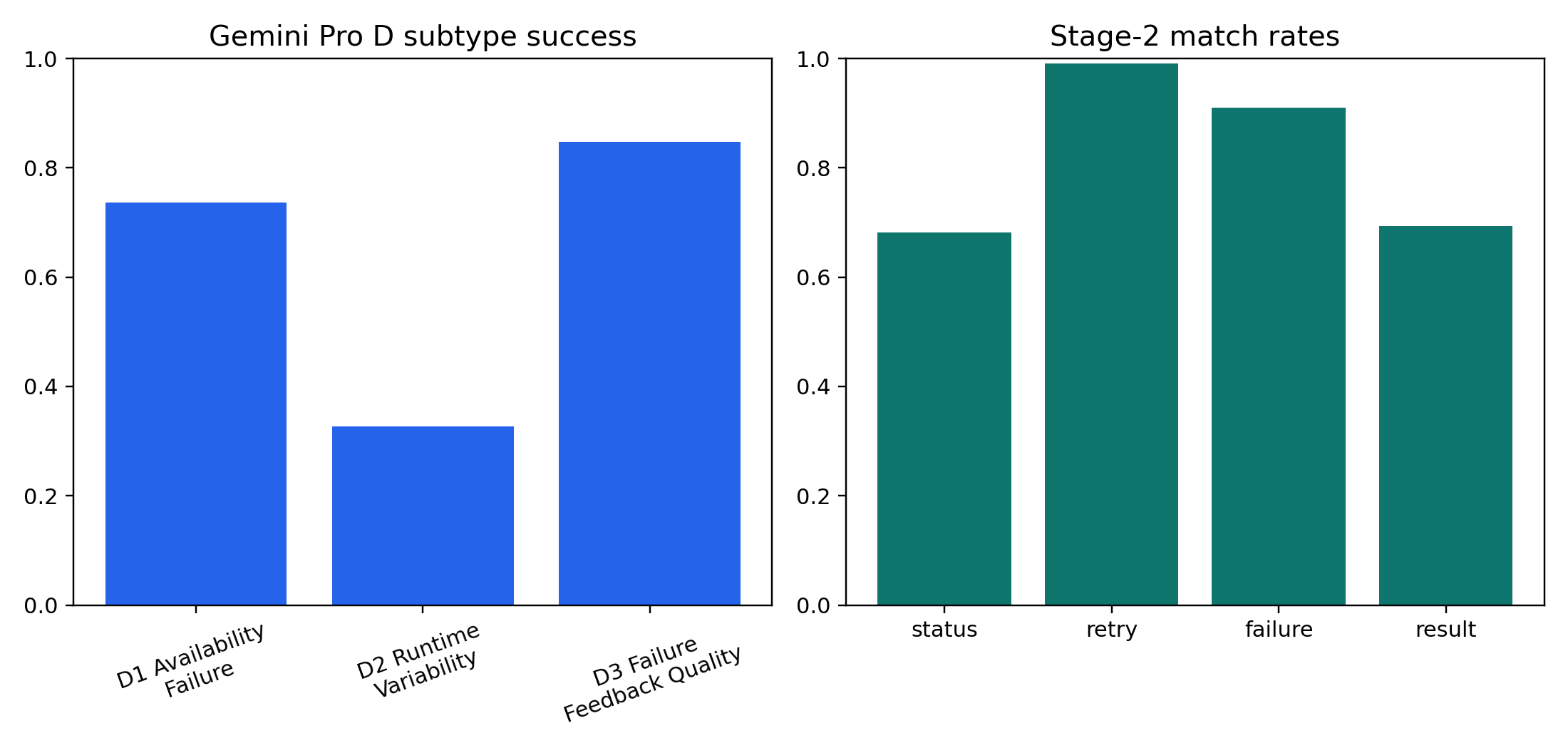}
\caption{Post-execution diagnostic profile for \texttt{gemini-2.5-pro}, separating status, retryability, failure-type, and final-result matching.}
\label{fig:app-gemini-profile}
\end{figure*}

\begin{figure*}[t]
\centering
\includegraphics[width=0.95\linewidth]{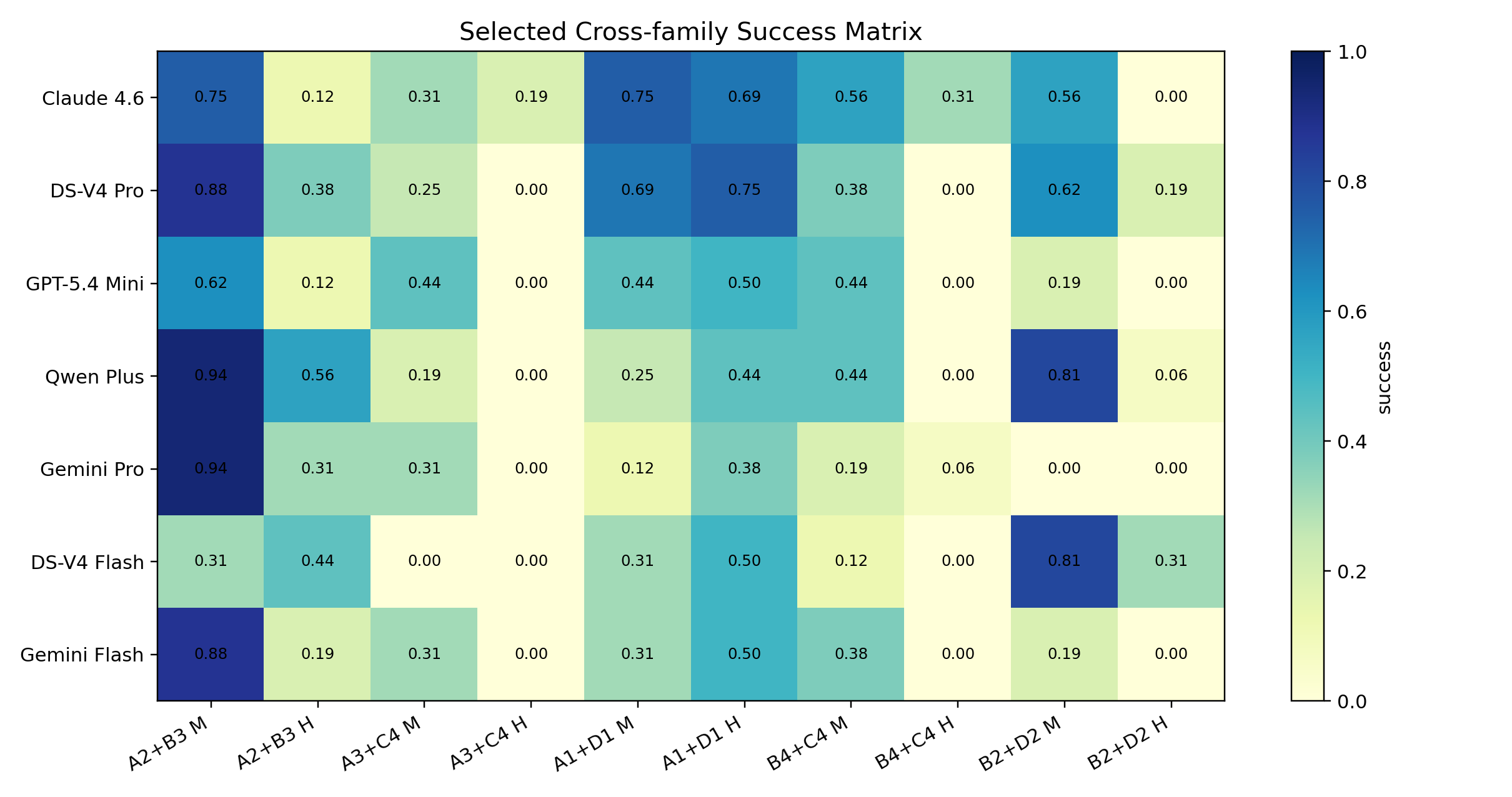}
\caption{Selected mixed-family success matrix for the five main-text subtype pairs at medium and heavy severity across evaluated models.}
\label{fig:app-cross-success}
\end{figure*}

\begin{figure*}[t]
\centering
\includegraphics[width=0.95\linewidth]{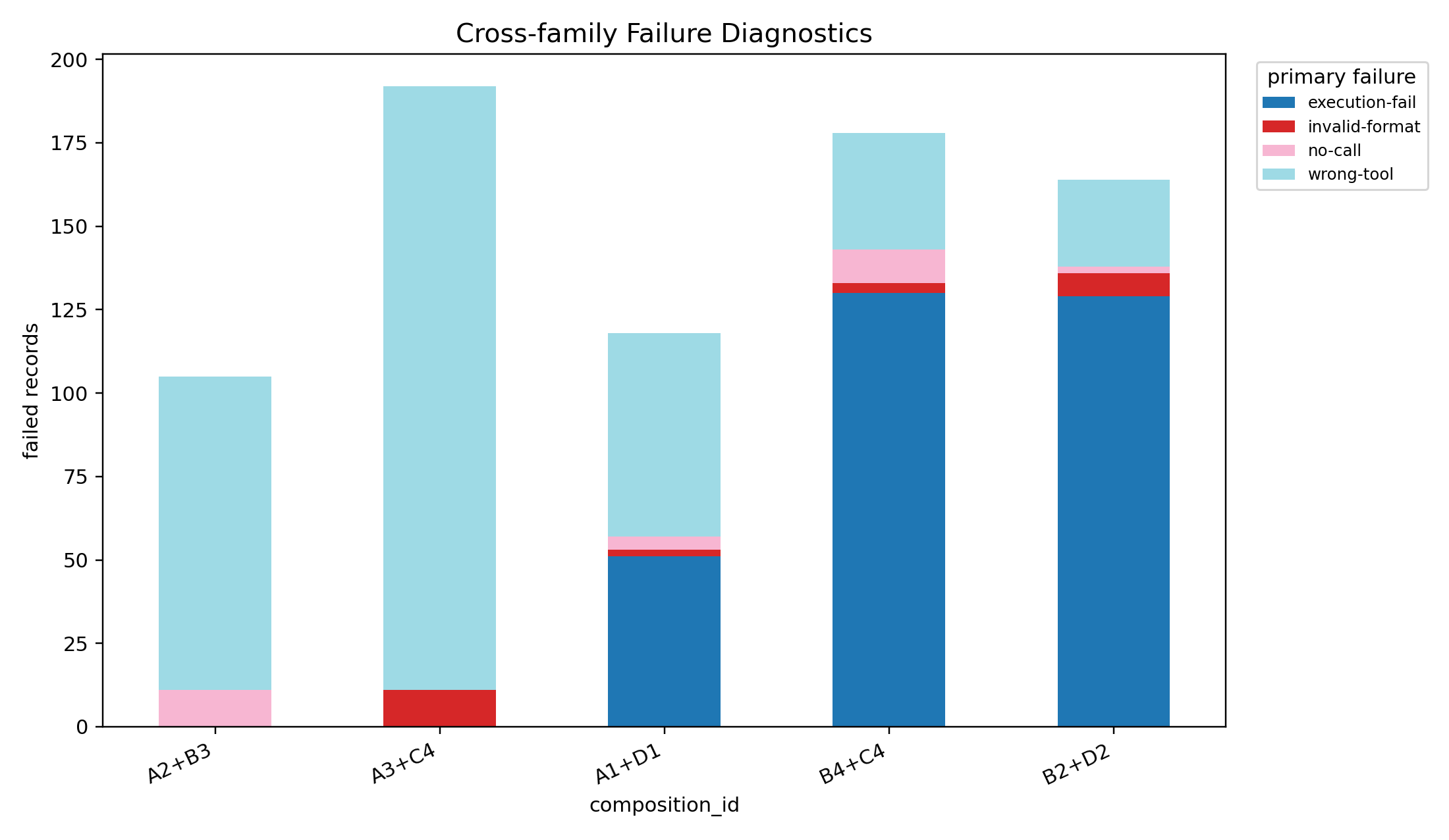}
\caption{Failure attribution for selected mixed-family pairs, reporting how interactions shift primary error types and cascade behavior.}
\label{fig:app-cross-failure}
\end{figure*}

\begin{figure*}[t]
\centering
\includegraphics[width=0.95\linewidth]{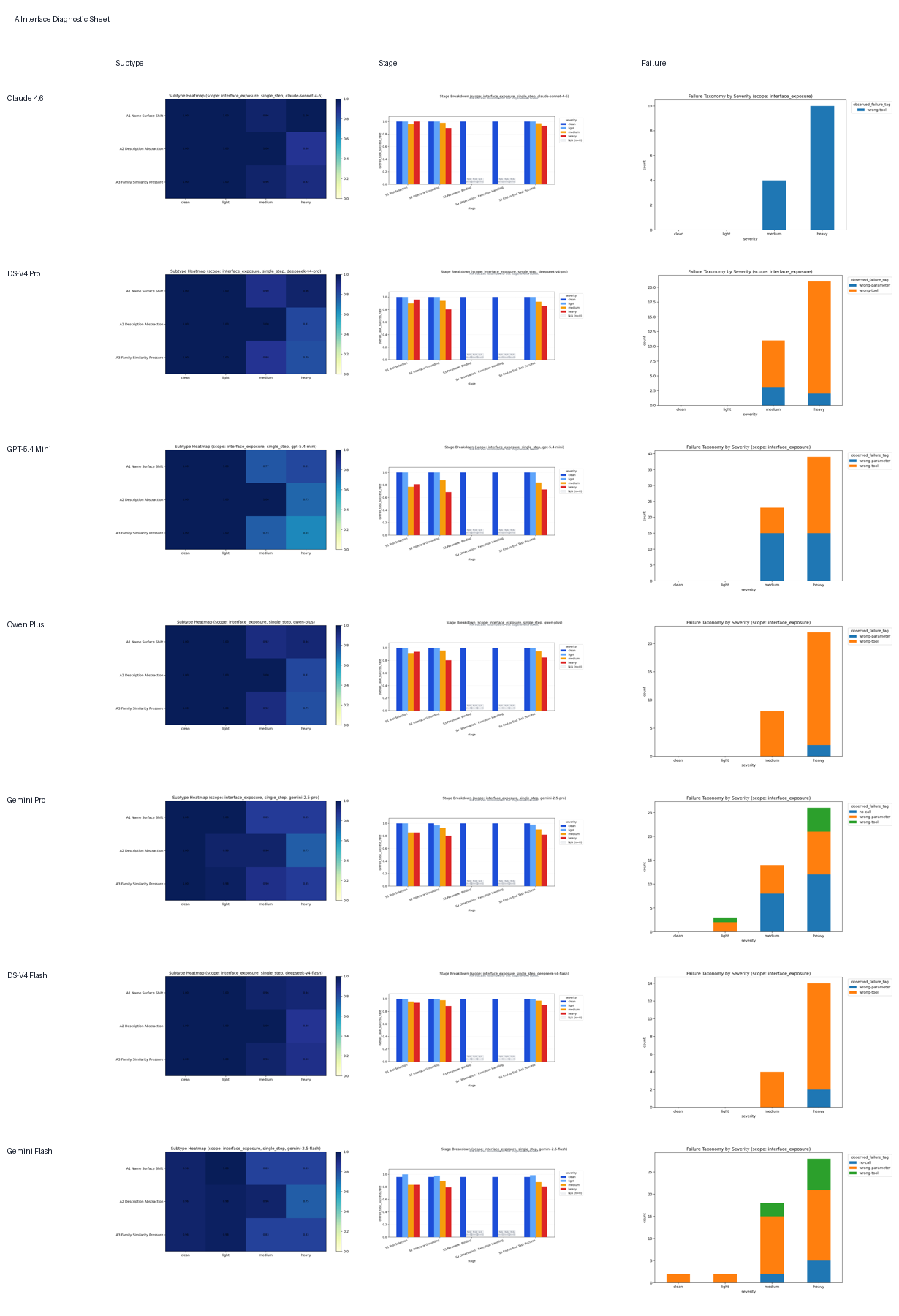}
\caption{Tool-interface diagnostic sheet for family A, including name shift, description abstraction, and family similarity pressure results.}
\label{fig:app-interface-sheet}
\end{figure*}

\begin{figure*}[t]
\centering
\includegraphics[width=0.95\linewidth]{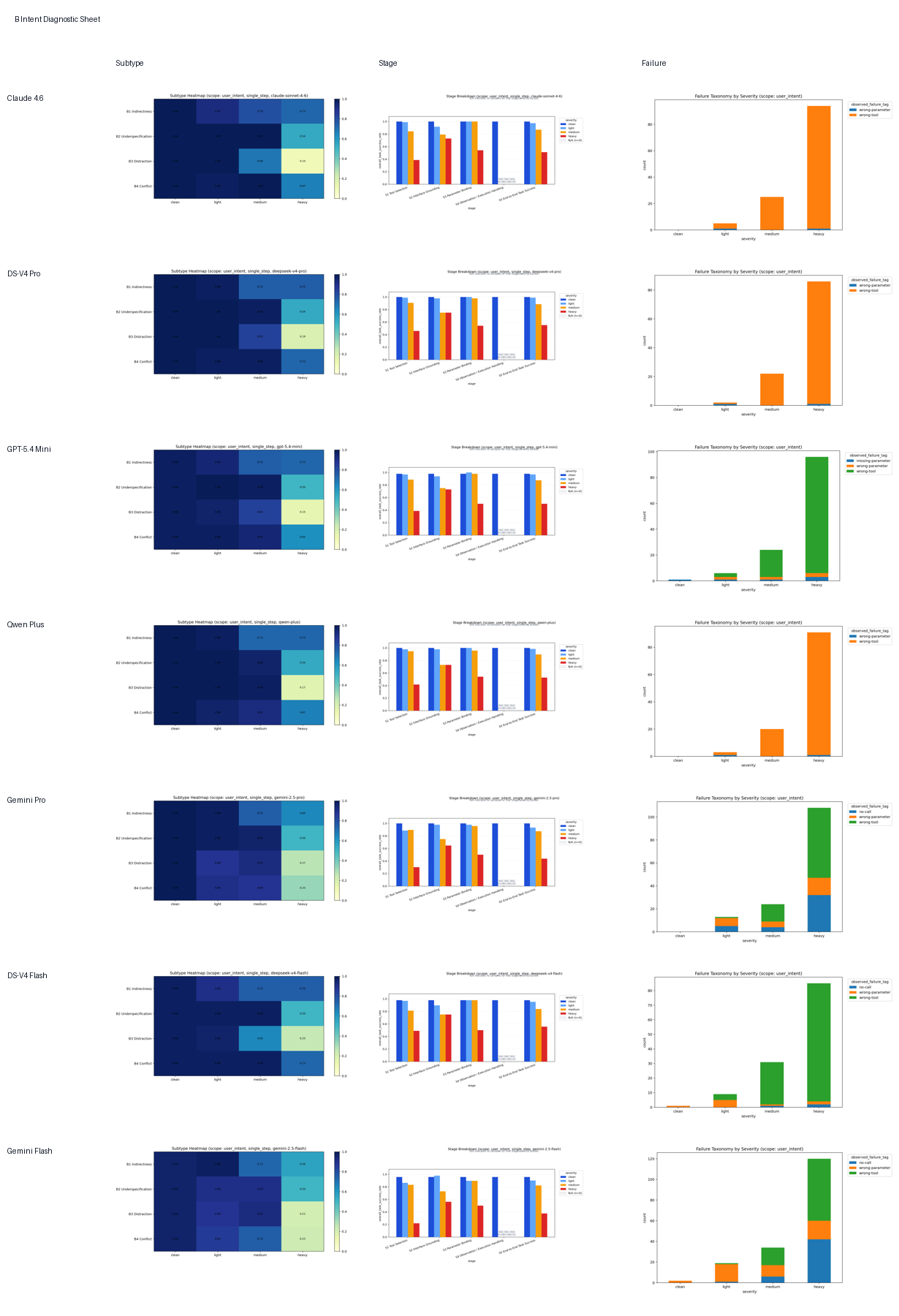}
\caption{User-intent diagnostic sheet for family B, including indirectness, underspecification, distraction, and conflict results.}
\label{fig:app-intent-sheet}
\end{figure*}

\begin{figure*}[t]
\centering
\includegraphics[width=0.95\linewidth]{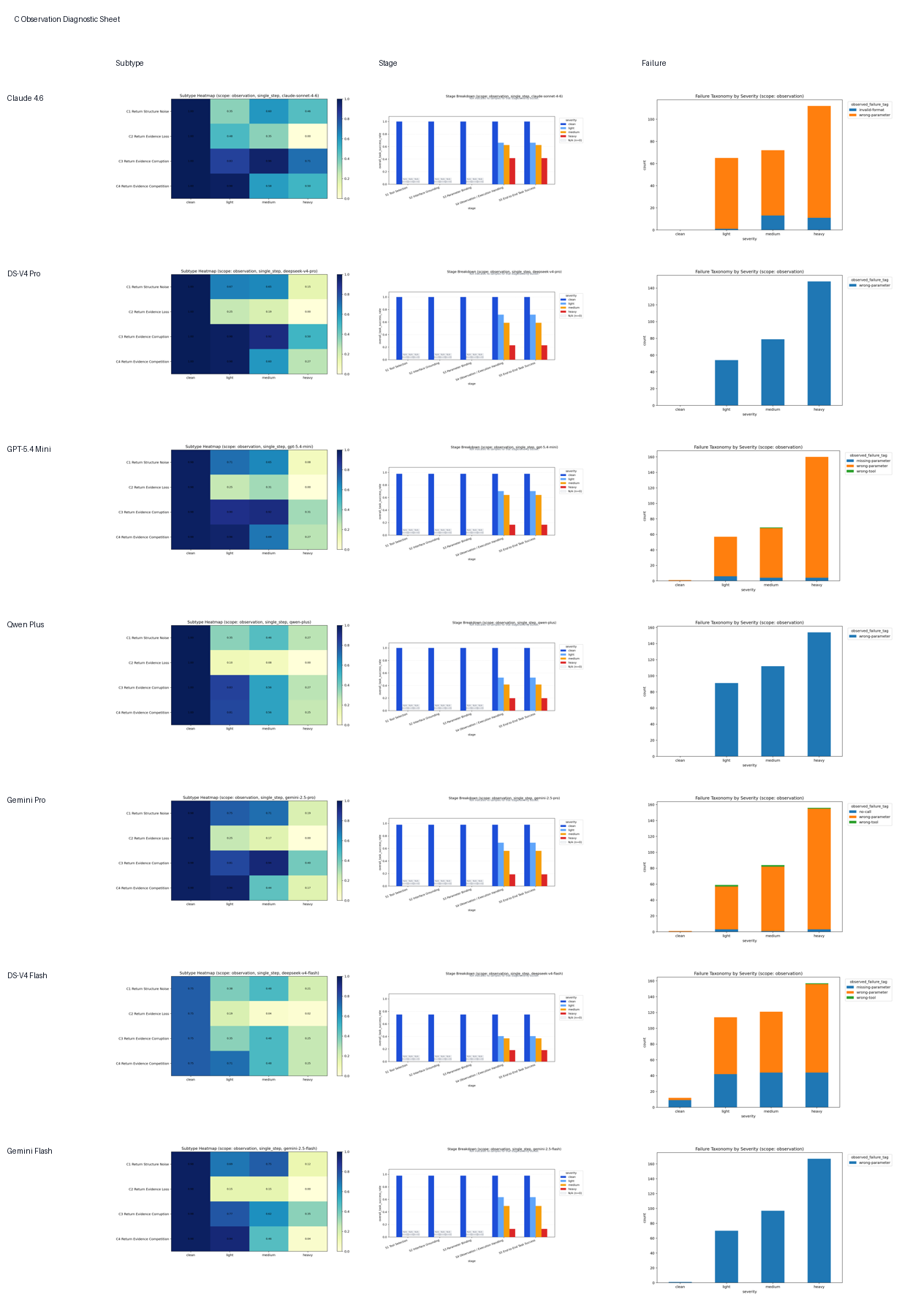}
\caption{Tool-output/observation diagnostic sheet for family C, covering return structure noise, evidence loss, corruption, and competition.}
\label{fig:app-observation-sheet}
\end{figure*}

\begin{figure*}[t]
\centering
\includegraphics[width=0.95\linewidth]{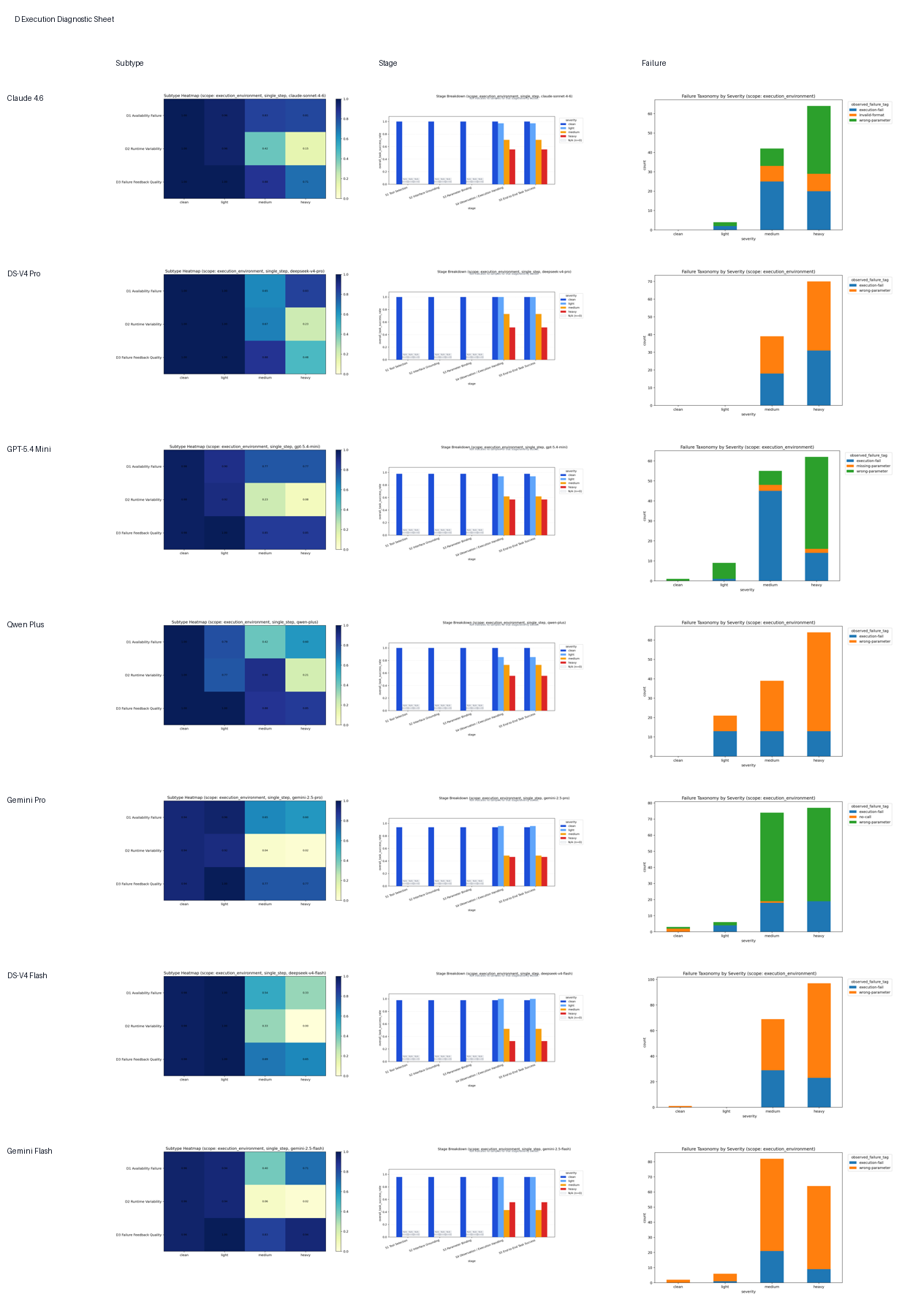}
\caption{Runtime-environment diagnostic sheet for family D, covering availability failure, runtime variability, and failure feedback quality.}
\label{fig:app-execution-sheet}
\end{figure*}

\begin{figure*}[t]
\centering
\includegraphics[width=0.95\linewidth]{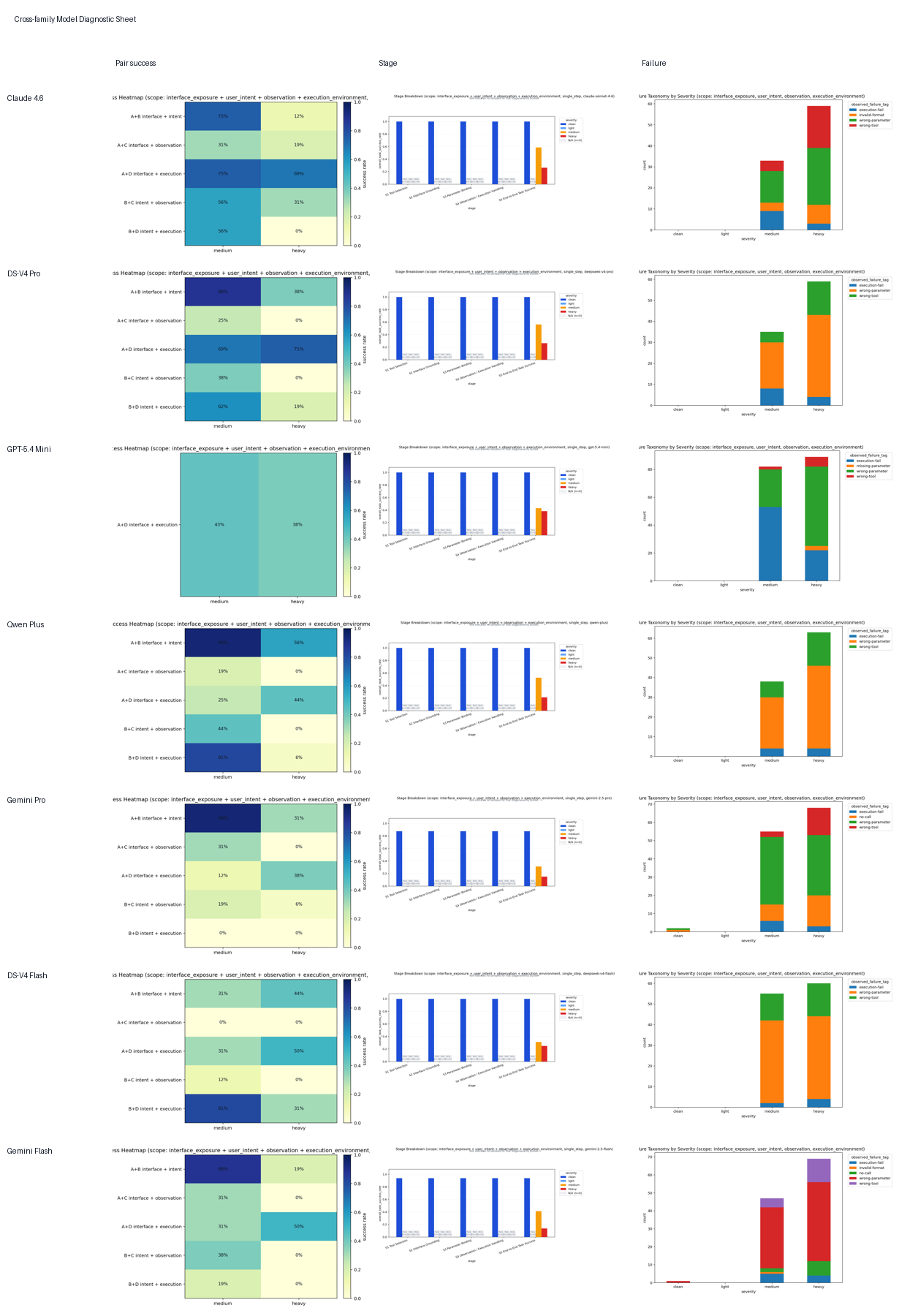}
\caption{Model-level summary of selected mixed-family interactions and their success/failure profiles.}
\label{fig:app-cross-model-sheet}
\end{figure*}

\begin{figure*}[t]
\centering
\includegraphics[width=0.95\linewidth]{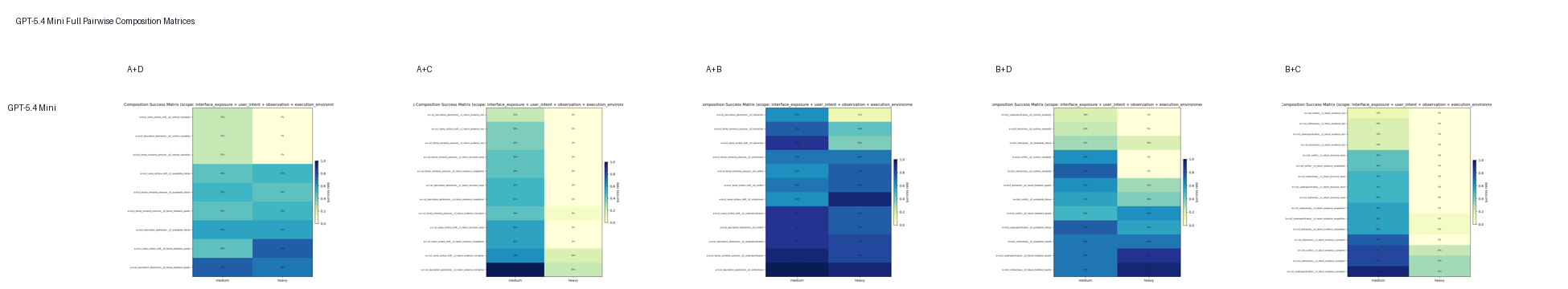}
\caption{Full mixed-family composition matrix for \texttt{gpt-5.4-mini}, retained to show the broader checked composition space beyond the five representative main-text pairs.}
\label{fig:app-gpt-cross-matrix}
\end{figure*}
\fi

\end{document}